\documentclass[twocolumn]{aastex701}
\usepackage{graphicx} 
\usepackage{subcaption}
\usepackage{amsmath}
\usepackage{longtable}

\begin{document}

\title{Magnesium silicate condensation in sub-Neptune envelopes: the fundamental link between chemistry, structure, and observables}

\correspondingauthor{William Misener}
\email{wmisener@carnegiescience.edu}

\author[0000-0001-6315-7118]{William Misener}
\affiliation{Earth and Planets Laboratory, Carnegie Institution for Science, 5241 Broad Branch Road NW, Washington, DC 20015, USA}
\email{wmisener@carnegiescience.edu}

\author[0000-0002-8518-9601]{Peter Gao}
\affiliation{Earth and Planets Laboratory, Carnegie Institution for Science, 5241 Broad Branch Road NW, Washington, DC 20015, USA}
\email{pgao@carnegiescience.edu}

\author[0000-0003-0354-0187]{Nicole L. Wallack}
\affiliation{Earth and Planets Laboratory, Carnegie Institution for Science, 5241 Broad Branch Road NW, Washington, DC 20015, USA}
\email{nwallack@carnegiescience.edu}

\author[0000-0002-6178-9055]{Maggie A. Thompson}
\affiliation{Earth and Planets Laboratory, Carnegie Institution for Science, 5241 Broad Branch Road NW, Washington, DC 20015, USA}
\email{mthompson@carnegiescience.edu}

\author[0000-0002-0794-2717]{Anat Shahar}
\affiliation{Earth and Planets Laboratory, Carnegie Institution for Science, 5241 Broad Branch Road NW, Washington, DC 20015, USA}
\email{ashahar@carnegiescience.edu}

\begin{abstract}
Chemical interactions between the hydrogen-dominated envelopes and silicate-rich interiors of sub-Neptunes likely play a key role in shaping their atmospheric structure, mass-radius relations, and upper atmosphere composition. While atmospheric abundances and structure deeply influence each other, many existing models have either considered the effects of chemical interactions without the structural implications or have modeled the envelope structure using oversimplified chemical networks. In this work, we introduce \texttt{Rocky Raccoon}, a coupled chemical equilibrium-atmospheric structure model. This model incorporates Mg, Si, O, C, and H species and produces self-consistent atmospheric chemical and thermal profiles for sub-Neptune envelopes, treating multi-species condensation for the first time. We find that the condensation sequence of magnesium silicates above a magma ocean is determined by the basal magma composition. We show that these condensation sequences drive the upper atmospheric composition to two endmembers: high oxygen abundances in the underlying melt produce compositions rich in oxygen-bearing volatiles (sub-solar C/O) and higher mean molecular weight atmospheres with $\mu \sim 4$~amu, while oxygen-poor melts produce lower mean molecular weight atmospheres dominated by methane and silane (super-solar C/O). The transition between the two regimes is abrupt and depends on melt properties like Mg/Si ratios and oxygen abundances. The different condensation sequences also lead to different thermal profiles, as deep convection is inhibited over different regions due to varying molecular weight gradients. Further experiments and simulations are key to resolving critical uncertainties in the condensation sequences and the corresponding significant impacts on sub-Neptune composition and thermal structure.
\end{abstract}

\section{Introduction} \label{sec:intro}
Sub-Neptunes, planets with radii between $\sim 2$ and 4 Earth radii and orbital periods within 100 days, typically have low bulk densities that can be reproduced by a degenerate range of mixtures of ices, rock, hydrogen, and vapor \citep{RS10}. Evolution models fit to observed demographics, such as the location of the radius valley \citep{Fulton17}, have been used to conclude that most sub-Neptunes have hydrogen-rich envelopes, with masses a few percent that of the planets, overlying Earth-like rock/iron interiors \citep{GS19, RO21}. These inferences on the structure of sub-Neptunes were initially made using simplified atmospheric and interior structure models assuming discrete planetary layers, typically fully adiabatic hydrogen-dominated envelopes that meet a silicate core \citep[e.g.,][]{LF14}. However, the community now broadly realizes that, at the temperatures and pressures expected at depth for much of a sub-Neptune's evolution, which can be $>2000$~K and $> 10000$~bar for gigayears respectively, the silicates at the base of the envelope ought to be molten \citep[e.g.,][]{CS18, Calder2026}, or even super-critical \citep[e.g.,][]{YoungStixrude2024, RogersYoung2025miscible, GilmoreStixrude2026}, allowing efficient chemical exchange between the atmosphere and interior. Such exchange could significantly alter the expected structure and chemistry of these planets.

Above the atmosphere-interior interface, this interchange is mediated by vaporized silicates, which are stable at equilibrium at these temperatures and pressures \citep[e.g.,][]{SY22}. These silicates recondense and rain out in cooler, lower pressure regions, removing heavy species from the gas and thus inducing a mean molecular weight gradient. The molecular weight gradient can be strong enough to inhibit convection, leading to regions above the magma ocean surface in which energy transport is limited to radiation and conduction \citep{MisenerSchlichting2022, Markham22}. The same physical mechanism has long been proposed for the Solar System giant planets, where the condensable species is typically water \citep{G95, L17, Clement2024}. In these non-convective regions, temperature profiles are typically super-adiabatic for realistic opacities and conductivities, leading to hotter interiors for the same surface conditions or, equivalently, a smaller observed radius for the same envelope mass and base temperature \citep{MisenerSchlichting2022}. Exactly how large these non-convective regions are and how long they persist depend on the chemical speciation of the melt and gas.

Complicating matters, the chemical composition of the envelope also changes due to the injection of evaporated, oxidizing silicates. Chemical equilibrium modeling predicts that deep within sub-Neptunes, silicate gas reacts with the background hydrogen, producing abundant water (H$_2$O) \citep[e.g.,][]{KiteBarnett2020, SY22} and silane (SiH$_4$) vapor \citep{MisenerSchlichting2023, CharnozFalco2023, Ito2025, HakimBower2026}. These predictions have been validated by high-pressure experiments \citep{Shinozaki14, MiozziShahar2025, HornVazan2025} and density functional theory models \citep{GilmoreStixrude2026}. Importantly, these chemical changes alter the mean molecular weight gradient itself, feeding back into the atmospheric structure calculations and producing different temperature and density profiles than a pure SiO vapor calculation. Thus, a fully coupled model of atmosphere-interior interactions is required to understand sub-Neptune evolution.

Understanding atmosphere-interior exchange has taken on new urgency with advancing \textit{JWST} observations of sub-Neptune upper atmospheres. Thus far, these observations have demonstrated that sub-Neptune atmospheric compositions vary significantly, ranging between near-solar \citep[$\mu$ $\sim$ 2 amu;][]{Davenport2025}, intermediate \citep[$\mu$ $\sim$ 5.5 amu;][]{BennekeRoy2024, BaratFairnington2026}, and water-like mean molecular weights \citep[$\mu$ $\sim$ 10-18 amu;][]{PiauletGhorayeb2024} and with various inferred hazes or clouds \citep{GaoPiette2023}. Interpreting these observations requires knowing how sub-Neptune interiors control upper atmosphere abundances, again necessitating a coupled model.  

Previously, studies have largely described the important gas-condensate chemistry only in pieces, with works typically simplifying either the chemical profiles or the envelope structure. In perhaps the simplest approximation, some models calculate atmospheric compositions at the envelope base conditions and extrapolate them to the upper atmosphere without including condensation at all, i.e. a `0-dimensional box' model of the atmosphere \citep[e.g.,][]{CharnozFalco2023, HengOwen2025, WerlenBurn2026}, while others employ chemical reactions but ignore silicate gas-phase chemistry entirely \citep[e.g.,][]{Rigby2024, ShorttleJordan2024, ItoChangeat2026}. Many works calculate the chemical rainout of silicates without self-consistently incorporating the structural changes these species cause \citep[e.g.,][]{LeeWerlen2025, Nixon2025, Ito2025, HakimBower2026, Steinmeyer2026, MukherjeeNixon2026}. Finally, works that do consider the structural effects of silicate rainout often employ an over-simplified set of condensates, often solely SiO$_2$ or MgSiO$_3$, and/or pared-down gas species networks \citep{MisenerSchlichting2022, Markham22, MisenerSchlichting2023, YoungStixrude2024, TejadaArevaloGupta2026}. These simplifications leave previous models unable to self-consistently connect planetary radii, melt composition, and atmospheric chemistry at once.

In reality, the entire suite of possible chemical interactions between silicates and more volatile species can lead to a wide variety of condensation sequences in sub-Neptune envelopes, and thus their thermal structure and upper atmosphere abundances. The power of Mg-Si condensates to sequester oxygen and other species has been previously shown to be important in super-Earth \citep{Herbort2022}, hot Jupiter \citep{LeeDobbs-Dixon2016} and brown dwarf \citep{lodders2002, Calamari2024} contexts. Given that sub-Neptunes are thought to be dominated by silicates in total mass, their interiors have the ability to strongly control the amount of oxygen in the envelope, thus dictating atmospheric composition and evolution.

In this work, we build on \citet{MisenerSchlichting2023}, which included self-consistent chemical-structural coupling but assumed that the magma was pure SiO$_2$, leaving silicon and oxygen as the only condensable elements and implicitly fixing their ratio, which other works have followed \citep{Ito2025, HakimBower2026}. However, realistic Earth-composition melts contain many other elements, most abundantly magnesium \citep[e.g.,][]{Palme2003}, which could be critical in controlling envelope mean molecular weight and composition. Meanwhile, atmospheric observations of sub-Neptunes have underscored the importance of carbon in their upper atmospheric chemistry \citep[e.g.,][]{Madhusudhan2023, BennekeRoy2024, RigbyMadhusudhan2025}, and sub-Neptune carbon abundances and C/O ratios have been invoked as evidence of atmosphere-interior interactions \citep{ShorttleJordan2024, Werlen2025C-O} and formation locations \citep[e.g.,][]{YangHu2024, Steinmeyer2026}. Therefore, including elemental species beyond Si, O, and H is critical for predicting and interpreting sub-Neptune observations.

The limitations in speciation in \citet{MisenerSchlichting2023} were imposed by an inflexible chemical framework optimized for the three element system considered there. In this work, we overcome these constraints by coupling a large chemical network, FASTCHEM COND \citep{FastchemOriginal, Fastchem2, FastchemCond}, with a self-consistent envelope structure model. We use the physical principles of condensation-induced molecular weight gradients described in \citet{MisenerSchlichting2022} and \citet{MisenerSchlichting2023}, enhancing their methods to incorporate multi-species condensation in a five element system including H, Mg, Si, O, and C. This coupled code, which we christen \texttt{Rocky Raccoon} due to its ability to model Rock vapor and Radial Atmospheric Composition, allows us to make predictions for how the interior abundance of key elements dictates the observable atmospheric compositions and planetary radii of sub-Neptunes. In doing so, we demonstrate the importance of coupled chemical-structural models in understanding exoplanet atmospheres, uncover the critical gaps in experimental data that can affect how we interpret observations, and make novel constraints on the bulk abundances of the most common planets yet discovered.

\section{Methods} \label{sec:methods}
In this section we describe our coupled chemistry-structure calculations. At a basic level, our procedure begins by calculating equilibrium chemical abundances at the temperature and pressure at the base of the envelope using FASTCHEM COND \citep{FastchemCond}. We then step down in pressure using a dynamic pressure increment chosen to be small relative to the local pressure $P$, $\Delta P=P/100$. We use the equations in Section \ref{sec:structure} to calculate the corresponding temperature and radius changes over this pressure step. At 
the new pressure, temperature, and radius, we calculate chemical equilibrium again, removing whatever has condensed in the previous step from the composition. We proceed as such until we reach the outer boundary of the atmosphere, which we choose to be $10^{-3}$~bar to capture the region observable with transmission spectroscopy. After doing so, we calculate the total mass, transit radius, and outer temperature of the obtained envelope profile. As the mass, radius, and outer temperature are not known \textit{a priori}, we iteratively calculate atmospheric profiles until we converge on a solution that matches the desired values. In the following subsections, we describe the chemical and structural methods in more detail.

\subsection{Chemistry}
For a given pressure, temperature, and elemental abundance, we use FASTCHEM COND \citep{FastchemCond} to calculate the equilibrium chemical composition. FASTCHEM COND is an open-source Gibbs energy minimization code that handles multi-species condensation appropriate for planetary atmospheres. We consider all species in the FASTCHEM COND library of condensates and gases that contain H, Si, and O in our most basic model, and then extend to those containing C and Mg. We adopt this simplified network of the most abundant elements to demonstrate the key physics in the system, at the expense of completeness. However, our library of condensates is more diverse than existing sub-Neptune condensate studies \citep{LeeWerlen2025}. Since the temperatures we consider in our modeling ($\sim$4000 K) are beyond the validity range of the Gibbs energies of many of the chemical species we treat, we are forced to extrapolate by setting FASTCHEM COND's \texttt{useCondDataValidityLimit} parameter to \texttt{False}. As we will show, uncertainties in the composition of condensates at the temperatures and pressures we consider affect our basic results,  thus underscoring the necessity of further chemical modeling and experiments. We exclude highly volatile species, e.g. CH$_4$(s), from our extrapolation when the temperatures and pressures are beyond the triple point of the species, as including them introduces spurious condensation in physically impossible high T/P regimes. We present the complete list of species we consider in Appendix~\ref{sec:appendix_tables}.

We initialize the model with elemental abundances in prescribed ratios that correspond to the planetary bulk composition, as described in Section~\ref{sec:results} below. FASTCHEM COND takes the temperature, pressure, and abundances and calculates the equilibrium gas and condensate number densities. While FASTCHEM COND is able to natively simulate rainout chemistry along an input pressure-temperature grid, we do not \textit{a priori} know the full self-consistent pressure-temperature profile. Therefore, we simulate rainout chemistry by first using FASTCHEM COND in equilibrium chemistry mode at each pressure level. We then subtract the elemental abundances of the equilibrium condensate from the overall elemental composition and use the new composition at the next pressure level, repeating this process upwards through the envelope.

\subsection{Structure}\label{sec:structure}
Our atmospheric structure model builds on \citet{MisenerSchlichting2022} and \citet{MisenerSchlichting2023}, which introduced the problem of mean molecular weight gradients due to silicate condensation in sub-Neptune envelopes. We extend the concepts introduced in those works by considering systems with multiple condensates and condensable gases acting at once. In this situation, the previously-formulated `Guillot criterion', the critical mass mixing ratio of condensable gas \citep[e.g.,][]{G95, L17, Markham22, MisenerSchlichting2022}, no longer applies. Instead, we calculate whether convection occurs at every pressure level by directly comparing the temperature and mean molecular weight gradients for the convective and non-convective profiles, similar to the \citet{L47} criterion but accounting for condensation in both. The atmosphere is convective if the density gradient for an adiabatic profile is smaller than that of a radiative profile, i.e.:
\begin{equation}\label{eq:ledoux}
    \left(\frac{\partial \ln T}{\partial \ln P}-\frac{\partial \ln \mu}{\partial \ln P}\right)_\mathrm{conv} <  \left(\frac{\partial \ln T}{\partial \ln P}-\frac{\partial \ln \mu}{\partial \ln P}\right)_\mathrm{non-conv},
\end{equation}
where $T$, $P$, and $\mu$ are the local temperature, pressure, and mean molecular weight respectively. Otherwise, the atmosphere is stable against convection. When the gradients in molecular weight are small, this equation reduces to the Schwarzschild criterion; such a situation occurs in the upper envelopes of our models and determines the upper radiative-convective boundary. In the case of a single condensable, this formulation reduces to the `Guillot criterion' \citep{G95, L17, MisenerSchlichting2022}.

To obtain the convective values, we calculate the adiabatic gradient for the gas mixture:
\begin{equation}\label{eq:adtemp}
    \left.\frac{\partial \ln T}{\partial \ln P} \right\rvert_\mathrm{conv} = \frac{k_\mathrm{B}}{\mu c_\mathrm{p}}
\end{equation}
where $k_\mathrm{B}$ is the Boltzmann constant and $c_\mathrm{p}$ is the local specific heat capacity. The latter value is calculated based on the FASTCHEM COND chemical equilibrium calculations, using a mass-weighted sum of the heat capacity of each constituent species $i$:
\begin{equation}
    c_{\mathrm{p}, i} = \frac{k_\mathrm{B}}{\mu_i}\frac{\gamma_i}{\gamma_i-1}.
\end{equation}
where $\mu_i$ and $\gamma_i$ are the molecular weight and adiabatic index of species $i$, respectively, with the latter assumed to be 1.3 for SiH$_4$ following that of the similarly structured methane; 4/3 for H$_2$O; and 7/5 for all other species, appropriate for diatomic molecules \citep{MisenerSchlichting2023}. As we will see, most other species are minor constituents in the envelopes we condsider, such that their mass-weighted contributions to the overall lapse rate are negligible. We assume a dry adiabat, neglecting the latent heat of vaporization, as \citet{MisenerSchlichting2023} demonstrated that it is virtually indistinguishable from the full moist adiabatic profile.

For the non-convective temperature gradients, we take the thermal diffusion equation:
\begin{equation}\label{eq:radtemp}
    \left.\frac{\partial \ln T}{\partial \ln P}\right\rvert_\mathrm{non-conv} = \frac{3 \kappa_\mathrm{eff} P L}{64\pi G M_\mathrm{c} \sigma T^4},
\end{equation}
where $G$ is the gravitational constant, $L$ is the planetary heat flow, $M_\mathrm{c}$ is the planet mass, and $\sigma$ is the Stefan-Boltzmann constant. $\kappa_\mathrm{eff}$ is an effective opacity which incorporates both conduction and radiation, which may be competitive at the conditions of the base of sub-Neptune envelopes \citep{Vazan2020, MisenerSchlichting2023, TejadaArevaloGupta2026}. We calculate $\kappa_\mathrm{eff}$ by combining the contributions of radiation and conduction to energy transport:
\begin{equation}
    \kappa_\mathrm{eff} = \cfrac{1}{\cfrac{1}{\kappa} + \cfrac{1}{\kappa_\mathrm{cond}}}
\end{equation}
where $\kappa$ is the Rosseland mean opacity and $\kappa_\mathrm{cond}$, the ``conductive opacity'', is given by recasting the thermal conductivity $\lambda_\mathrm{cond}$ as
\begin{equation}
    \kappa_\mathrm{cond} = \frac{16 \sigma T^3}{3 \rho \lambda_\mathrm{cond}}.
\end{equation}
The material properties of the high-pressure, exotic mixtures of silicates and hydrogen we consider here, including their thermal conductivities, opacities, and non-ideal behavior, are highly uncertain. We employ a simple approach here, following the methods of \citet{MisenerSchlichting2023}, using thermal conductivities based on hydrogen experiments \citep{McWilliams2016} and opacities appropriate for solar composition mixtures \citep{F14}. This estimate of the opacities likely provides a lower bound, as \citet{F14} omitted high-temperature short-wave absorbers such as metal hydrides that act to further increase the Rosseland mean opacities \citep{Siebenaler2026}. As in \citet{MisenerSchlichting2023}, we find here that conduction dominates radiation at depth, a result that would only become more entrenched for higher opacities. However, the thermal conductivities at these temperatures, pressures, and compositions are uncertain and the subject of ongoing research \citep[e.g.,][]{Karasiev2026}. Uncertainties in the assumed conductivities can lead to large differences in the envelope structures obtained \citep{EberleinHelled2026}. 

We do not model wavelength-dependent opacities, and therefore obtain an isothermal upper atmosphere at the equilibrium temperature. Obtaining the true thermal profile using wavelength-dependent opacities set by the chemical composition of the envelope, including both gases \citep[e.g.,][]{PietteGao2023} and condensates \citep{MukherjeeNixon2026}, would be interesting future work, as it can feed back on the upper envelope condensate profile and affect atmospheric escape and observables \citep{MisenerSchulik2025,JanssenMiguel2026}.

Equation \ref{eq:radtemp} depends on the local pressure, temperature, effective opacity, and local energy flux $L$. We assume that, in steady state, the energy flux that must be transported across the radiative boundary is equal to the radiative flux of the planet to space, i.e., the luminosity at the upper radiative-convective boundary:
\begin{equation}\label{eq:luminosity}
    L = \frac{\gamma_\mathrm{rcb}-1}{\gamma_\mathrm{rcb}}\frac{64\pi G M_\mathrm{c} \sigma T_\mathrm{rcb}^4}{3 \kappa_\mathrm{rcb} P_\mathrm{rcb}},
\end{equation}
where `rcb' subscripts represent values calculated at the conditions of the upper radiative-convective boundary and $\gamma_\mathrm{rcb}$ is, similarly to $c_\mathrm{p}$, a mass weighted average. We discuss whether this assumption holds for all sub-Neptunes in Section~\ref{sec:discussion}. 

In all regimes, the radial pressure gradient at a radius $R$ is found following hydrostatic equilibrium:
\begin{equation}\label{eq:dPdr}
    \frac{\partial P}{\partial R} = -\frac{G M_\mathrm{c}}{R^2} \frac{\mu P}{k_\mathrm{B} T}.
\end{equation}

\noindent At every pressure step $\Delta P$, we determine whether the atmosphere is in a convective or non-convective regime by calculating the local convective and non-convective pressure gradients via Eqs.~\ref{eq:adtemp} and \ref{eq:radtemp}. We calculate the appropriate temperature steps for each regime $r$: $T_{\mathrm{new}, r} = T + \Delta P (\partial T/\partial P)_r$. To calculate the molecular weight gradients, i.e. the second term on each side of Eq.~\ref{eq:ledoux}, we use the new temperatures $T_{\mathrm{new}, r}$ to recalculate new chemical equilibria and therefore new corresponding mean molecular weights. We then compute the gradient in molecular weight and therefore directly calculate the stability to convection via Eq.~\ref{eq:ledoux}.

\subsection{Building the envelope model}
To compute the structure model, we begin at the base of the envelope, which we define as a `core' radius appropriate for silicate interiors: $(R_\mathrm{c}/R_\oplus) = (M_\mathrm{c}/M_\oplus)^{1/4}$ \citep[e.g.,][]{Valencia06}. In reality, the base of the envelope may be set by the onset of complete miscibility between silicates and hydrogen \citep{YoungStixrude2024, RogersYoung2025miscible, GilmoreStixrude2026}: we discuss how to incorporate this facet in future work in Sec.~\ref{sec:discussion}. We then choose a base temperature, which effectively corresponds to a planetary entropy. In this work, we choose to standardize to a base temperature of 4000~K, an appropriate choice for sub-Neptunes with substantial envelopes, which have large initial energy budgets from their binding energies \citep[e.g.,][]{GSS16, Vazan2018, MisenerSchlichting2022, TejadaArevaloGupta2026}, but in reality this temperature evolves in time as the planet cools. However, the base pressure $P_\mathrm{base}$ and energy flow through the planet $L$ remain unknown. To constrain these and close the system numerically, we choose an outer temperature $T_\mathrm{eq}$ and an envelope hydrogen mass $f$, expressed as a fraction of the planet's total mass. This mass includes hydrogen that has reacted with other species in the envelope, but not hydrogen that has ingassed to the silicate interior, which may host a substantial fraction of the hydrogen reservoir in sub-Neptunes \citep{CS18, RogersYoung2025miscible, HakimBower2026}. 

We solve for the atmospheric structure, defined by $P_\mathrm{base}$ and $L$, by employing the radiative-convective calculations discussed in Sec. \ref{sec:structure} that fits these two constraints using \texttt{scipy}'s \texttt{fsolve} method \citep{scipy} with a fractional tolerance of $10^{-8}$ in the resultant $f$ and $T_\mathrm{eq}$. As we will show, this calculation typically produces an atmospheric structure with (1) one or more deep non-convective regions produced by molecular weight gradient induced by silicate condensation, (2) one or more convective regions, and (3) an upper radiative region caused by satisfaction of the constant-$\mu$ Schwarzschild criterion at and above $R_\mathrm{rcb}$. We assume that the transit radius $R_\mathrm{t}$ corresponds to that at the 20~mbar pressure level \citep[e.g.,][]{LF14}. 

\section{Results} \label{sec:results}
\subsection{Si-O-H system}
\begin{figure*}
    \centering
    \includegraphics[width=\linewidth]{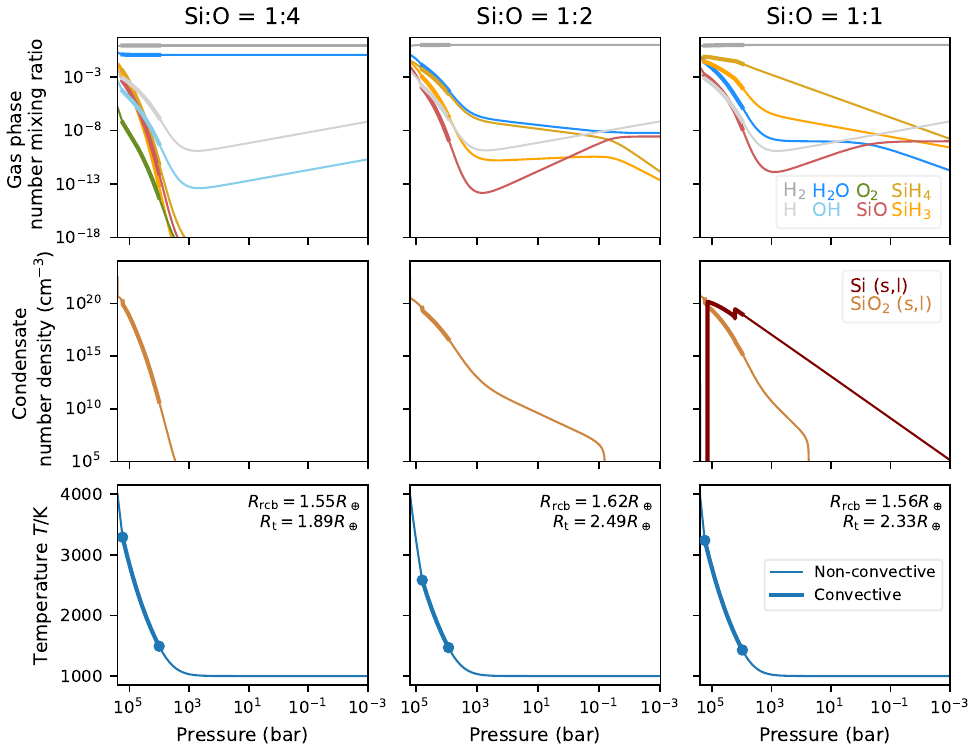}
    \caption{Comparison of select atmospheric gas abundances (top row), condensates (middle row), and temperature profiles (bottom row) for basal melt Si:O ratios of 1:4 (left), 1:2 (center), and 1:1 (right) in a sub-Neptune with $M_\mathrm{c}=4 M_\oplus$, $T_\mathrm{eq} = 1000$~K, $T_\mathrm{base}=4000$~K, and $f=0.03$. Thick lines show convective regions and thin lines show non-convective regions; dots on the bottom panel mark the transitions. The outer RCB and transit radii of the three models are given in the bottom panels. The Si:O=1:2 case (middle column) has conditions comparable to the models shown in \citet{MisenerSchlichting2023} and reproduces their results well. The 1:4 Si:O case (left column) has a higher oxygen abundance, leading to rapid silicon depletion and a water-rich atmosphere. Conversely, in the 1:1 Si:O case (right column), a relatively high Si:O ratio, oxygen is depleted at depth and SiH$_4$ remains abundant into the upper atmosphere.}
    \label{fig:SiOH_comp}
\end{figure*}
To demonstrate the possible variations in atmospheric structure, we begin by expanding on the Si-O-H system analyzed in \citet{MisenerSchlichting2023}. We find that varying the Si:O ratio in the basal melt below the envelope can lead to drastic changes in the atmospheric abundance profiles, temperature structure, planet radius (Figure~\ref{fig:SiOH_comp}). For a case with a 1:2 Si:O ratio at the base, similar to that studied in \citet{MisenerSchlichting2023}, we find SiH$_4$ and H$_2$O are the dominant secondary species to low pressures in equal, sub-solar composition, in line with \citet{MisenerSchlichting2023} (Figure~\ref{fig:SiOH_comp}, middle column). The model now allows for other gas species such as SiH$_3$ and H; here and in plots to follow, we show only the most relevant subset of all species present in the equilibrium calculation, the full suite of which are listed in Appendix~\ref{sec:appendix_tables}. We find in this case that these additional species play only minor roles at chemical equilibrium, validating the limited chemical network approximation of \citet{MisenerSchlichting2023} and agreeing with other recent chemical models \citep{Ito2025}. We also find that a non-convective region exists between the base of the envelope and where the temperature drops to $\sim$2600~K, similar to \citet{MisenerSchlichting2023}. The model presented here improves on previous work by self-consistently calculating the inner and outer convection criteria, allowing the outer radiative-convective boundary to be reached at $T>T_\mathrm{eq}$: here it is reached at $T \sim 1200$~K, $P \sim 10^4$~bar, and $R \sim 1.62 R_\oplus$. The transit radius is $2.49 R_\oplus$, squarely within the sub-Neptune regime, and the upper envelope mean molecular weight is 2.02~amu, reflecting the predominance of H$_2$.

Varying the Si:O ratio, not possible in the framework of \citet{MisenerSchlichting2023}, allows for novel atmospheric compositions and demonstrates the observable effects of atmosphere-interior interactions. In a silicon-depleted, oxygen-rich scenario where the Si:O ratio is 1:4, abundant water vapor forms at the base of the envelope, dwarfing the reduced and oxidized silicon-bearing gases (Fig.~\ref{fig:SiOH_comp}, left column). In this case, the silicon quickly rains out as SiO$_2$, leaving only water as the dominant secondary species after hydrogen. The remaining water significantly increases the mean molecular weight of the atmosphere to 3.88~amu and thus depresses the transit radius to $1.89 R_\oplus$, 30\% lower than in the 1:2 scenario despite equal atmospheric H masses. A significantly different atmospheric mass would be inferred from mass and radius alone for these two planets, underscoring the importance of considering interior structure.

Increasing the Si:O ratio also has marked effects on atmospheric abundances. For an oxygen-poor base melt composition, with a Si:O ratio of 1:1, SiH$_4$ is the dominant gaseous species, outproducing water at the base of the envelope (Fig.~\ref{fig:SiOH_comp}, right column). Here, rather than being Si-limited, the atmosphere becomes O-limited: the oxygen rains out as SiO$_2$, depleting the envelope of H$_2$O and leaving only reduced silicon species such as SiH$_4$ and SiH$_3$. These species then slowly rain out as atomic Si condensate, an oxygen- and carbon-poor silicon condensation species not typically considered \citep{Visscher2010, KitzmannStock2026}. The discontinuity in Si condensate number density visible at $\sim 10^4$~bar in Fig.~\ref{fig:SiOH_comp} is due to a discontinuity in the source Gibbs energy fits \citep{NIST}, highlighting that significant uncertainties exist in these data and may affect sub-Neptune models. This compositional endmember is similar to the cases of efficient water ingassing considered in \citet{Ito2025} and \citet{HakimBower2026}, which also produced low oxygen abundances in the atmosphere. However, \citet{Ito2025} neither included atomic Si condensation nor considered the effects of condensation on envelope structure. This Si-rich envelope leads to a radius detectably smaller than the 1:2 Si:O case, with a new transit radius of $2.34 R_\oplus$, along with a smaller non-convective region near the envelope base due to the fast condensation of silicate species that extends to only $\sim 3200$~K.

\subsection{Mg-Si-O-C-H system}
\begin{figure*}
    \centering
    \includegraphics[width=\linewidth]{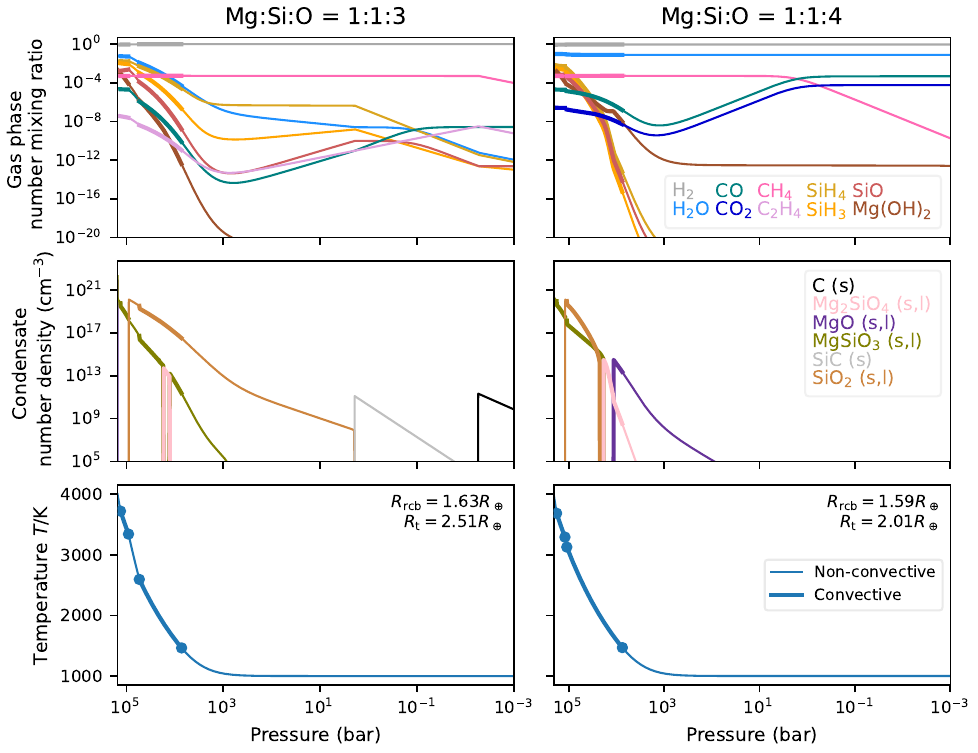}
    \caption{Comparison of atmospheric gas abundances (top), condensates (middle), and temperature profiles (bottom) for envelope base Mg:Si:O ratios of 1:1:3 (left) and 1:1:4 (right), for a sub-Neptune with $M=4 M_\oplus$, $T_\mathrm{eq} = 1000$~K, $T_\mathrm{base}=4000$~K, and $f=0.03$. As in Fig.~\ref{fig:SiOH_comp}, thick lines show convective regions and thin lines show non-convective regions; dots on the bottom panel mark the transitions. Increasing the oxygen abundance at the base alters the atmospheric radius by twenty percent and modifies the upper atmospheric chemistry from methane-silane dominated to water-carbon monoxide dominated at chemical equilibrium.}
    \label{fig:MgSi_comp}
\end{figure*}

We now extend our study to a five element system more relevant to realistic melt and atmosphere compositions. In addition to Si, O, and H, we add magnesium, abundant in rocky material, and carbon, an abundant volatile and key atmospheric observable. We begin by modeling a planet with equal bulk abundances of magnesium and silicon, appropriate for bulk silicate Earth and near-solar \citep{Palme2003}, and an envelope with solar carbon abundance \citep{Asplund2009}. We choose two magnesium:silicon:oxygen number ratios in the melt: 1:1:3, similar to bulk silicate Earth \citep{Palme2003}, to represent an oxygen-poor scenario, and 1:1:4 to represent an oxygen-rich scenario. We find that deep sub-Neptune envelopes can host a wide range of equilibrium condensate species, including those predicted for and observed on hot Jupiters and substellar objects, such as silicon oxides (SiO$_2$), magnesium silicates including forsterite (Mg$_2$SiO$_4$) and enstatite (MgSiO$_3$), and periclase (MgO) \citep{Visscher2010,Burningham21,grant2023,hoch2025}, as well as more unusual condensates such as silicon carbide (SiC) and atomic carbon (Figure~\ref{fig:MgSi_comp}). The wide range in abundance ratios, especially of oxygen, in these condensates means that when each species is condensing, it affects the remaining gas phase abundances differently. Therefore, self-consistent gas-condensate interactions produce intricate feedbacks between chemical equilibrium and condensation, and therefore between melt and atmospheric composition.

The variations in atmospheric state are exemplified by the impact of changing oxygen abundance at the base of the envelope. In the oxygen-poor case (1:1:3 Mg:Si:O ratio, Fig.~\ref{fig:MgSi_comp}, left column), the atmospheric oxygen nearly completely rains out at depth as SiO$_2$ and MgSiO$_3$, resulting in a strongly reducing, CH$_4$-SiH$_4$ upper atmosphere. In oxygen fugacity terms, using the scaling from \citet{Hirschmann2021}, the oxygen fugacity drops from $\Delta$IW$-4.5$ to $-23$ from the bottom to the top of the model. At pressures below a few bar, SiC begins to condense at equilibrium, depleting the remaining silicon. The quenching of this reaction therefore determines whether SiH$_4$ is detectable at pressures probed by transmission and emission observations. Low-pressure SiC condensation in very O-poor envelopes has been noted \citep{JanssenMiguel2026}, but not in such a Si-rich envelope. We also note the abundance of the carbon-carbon bonded hydrocarbon C$_2$H$_4$, with a 1~bar volume mixing ratio of $\sim 10^{-10}$ in the oxygen-poor case, which may be sufficient for efficient soot haze formation \citep{YangKempton2026}. Thus, oxygen condensation in sub-Neptune envelopes due to an O-poor silicate interior could favor aerosol formation.

In contrast to the oxygen-poor case, the atmospheric Si and Mg rains out quickly as a wide variety of condensates in the oxygen-rich scenario (1:1:4 Mg:Si:O ratio, Fig.~\ref{fig:MgSi_comp}, right column): first as MgSiO$_3$, then transitioning to SiO$_2$, Mg$_2$SiO$_4$, and MgO. These condensates deplete the atmosphere in gaseous Si and Mg species, leaving abundant oxygen behind and producing an H$_2$O-CH$_4$ dominated envelope at intermediate pressures. At lower pressures $\sim 1$~bar, the carbon-bearing species switches from CH$_4$ to CO and CO$_2$, as expected for our chosen 1000~K equilibrium temperature \citep{Moses2013}. The oxygen fugacity here increases from $\Delta$IW$-4.9$ to $-1.5$ through the atmosphere, demonstrating that an oxygen-rich upper atmosphere does not necessitate oxidizing conditions immediately above the magma ocean surface.

In both oxygen-poor and rich cases, consideration of condensation has amplified small differences in the oxygen abundance at the base of the envelope: a slightly lower oxygen abundance leads to complete oxygen depletion, while a slightly higher abundance produces copious atmospheric oxygen-bearing species. While similar effects have been considered for giant planets and brown dwarfs, where silicate cloud formation can reduce water vapor abundances by $\sim$10\% \citep{lodders2002}, here the impact is greatly amplified since most of the accreted oxygen for sub-Neptunes is in their silicate interiors, not their relatively low-mass envelopes. These results also show that condensation alters the oxygen fugacities in the envelope by orders of magnitude in either direction from the envelope base to the upper atmosphere depending on the condensate sequence.

\begin{figure}
    \centering
    \includegraphics[width=\linewidth]{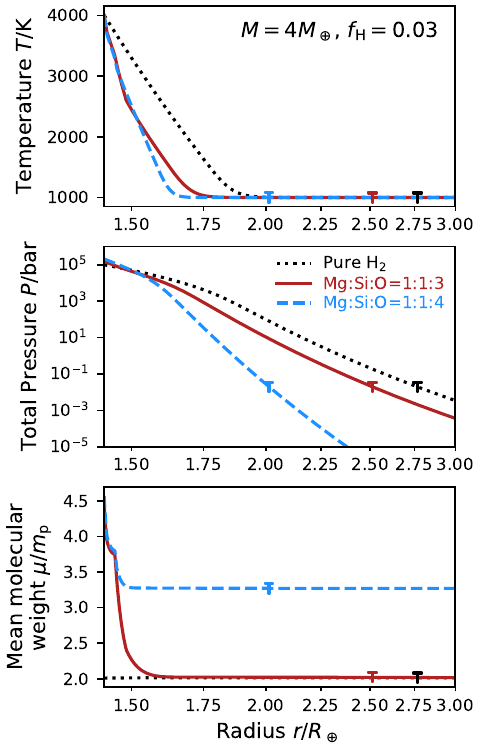}
    \caption{Comparison of radial profiles of temperature (top), pressure (middle), and mean molecular weight (bottom) for a pure H$_2$ model (black dotted) and models with silicate condensation and a Mg:Si:O ratio of 1:1:3 (red solid; same as the left column of Fig.~\ref{fig:MgSi_comp}) and 1:1:4 (blue dashed; same as the right column of Fig.~\ref{fig:MgSi_comp}). All models assume the same planet mass ($4 M_\oplus$), atmospheric hydrogen mass ($f=0.03$), temperature at the envelope base (4000~K), and equilibrium temperature (1000~K). `T' marks the transit radius. Both the non-convective regions at the base of the envelope and the remaining molecular species after condensation alter the transit radii and scale heights of sub-Neptunes.
    }
    \label{fig:radii}
\end{figure}

The chemical profiles we obtain result in different self-consistent atmospheric structures, as we show in Fig.~\ref{fig:MgSi_comp} and expand on in Figure~\ref{fig:radii} through comparison to a pure-H$_2$ envelope case more similar to typical mass-radius relations \citep[e.g.,][]{LF14}. Such an envelope is fully convective and has a constant, low mean molecular weight, $\mu \sim 2$~amu, and therefore extends the furthest, to a transit radius of $2.77 R_\oplus$, corresponding to a bulk density of $\sim 1.0$~g~cm$^{-3}$. Meanwhile, the oxygen-poor, Mg:Si:O=1:1:3 model possesses super-adiabatic regions at depth caused by condensation that steepens the slope of the temperature profile, decreasing the overall planet radius. After efficient rainout, its upper envelope is dominated by H$_2$, leading to a similar scale height as the pure H$_2$ (condensation-free) model. Here, the transit radius is $2.51 R_\oplus$, 10\% lower than the pure H$_2$ case, with a bulk density of $\sim 1.4$~g~cm$^{-3}$. Finally, the oxygen-rich model not only features super-adiabatic regions at depth caused by the molecular weight gradient from silicate condensation, its condensation sequence also produces significant water and CO that survives to the upper atmosphere, leading to a higher mean molecular weight, $\mu \sim 3.3$~amu, and a corresponding decrease in the scale height. These effects act together to decrease the radius significantly compared to the other cases, with a transit radius of $2.01 R_\oplus$ and a correspondingly larger density of $\sim 2.7$~g~cm$^{-3}$. 

We also note the formation of multiple deep non-convective regions, separated by detached convective zones, once multi-species condensation is included (Fig.~\ref{fig:MgSi_comp}). Transitions occur as the efficiency of condensation, dictated by the condensates stable at each pressure, temperature, and composition, changes as the gas composition evolves. Each condensate removes gas species at a different rate, altering the balance in Eq.~\ref{eq:ledoux}. Such sensitivity highlights the importance of self-consistent multi-species condensation in sub-Neptunes.

\begin{figure}
    \centering
    \includegraphics[width=\linewidth]{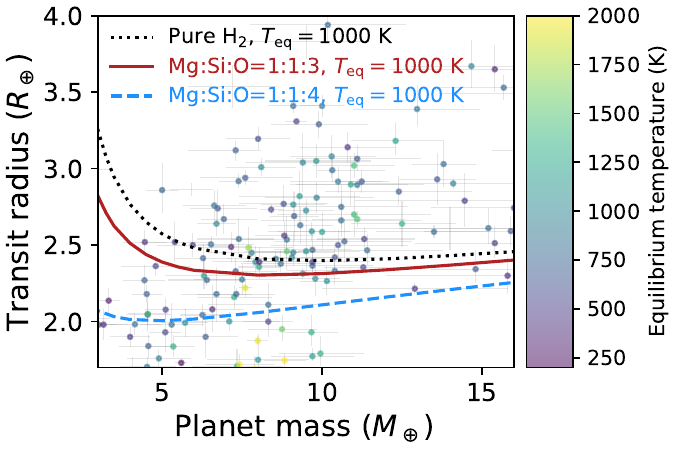}
    \caption{Mass-radius relationships for planets with a fixed envelope base temperature $T_\mathrm{base}=4000$~K, a fixed envelope hydrogen mass of $f_\mathrm{H}=0.03$, and a fixed equilibrium temperature of 1000~K. Line colors match those of Fig.~\ref{fig:radii}. The population of sub-Neptunes with precise masses ($\Delta M < 25\%$) and radii ($\Delta R < 10\%$) is shown for reference, taken from the NASA Exoplanet Archive \citep{Christiansen2025, PSCompPars}; they are colored by equilibrium temperatures calculated consistently using the host star's effective temperature, radius, and the planet semi-major axis. Inclusion of atmosphere-interior interactions shrinks radii across masses, most dramatically for small planets and when significant volatiles remain uncondensed.
    }
    \label{fig:MR_diagram}
\end{figure}

We construct mass-radius curves from our models assuming a fixed envelope base temperature of 4000~K, an envelope hydrogen mass of 0.03 planet masses, and an equilibrium temperature of 1000~K (Figure~\ref{fig:MR_diagram}). Across the sub-Neptune mass range, we find that including non-convective regions caused by condensation shrinks planet radii by $\geq 10$\%. The curves exhibit a strong dependence on the envelope base chemistry: oxygen-rich models are significantly smaller due to their higher atmospheric mean molecular weights. The spread in observed exoplanet radii for a given mass is comparable to the spread in our models, suggesting that some of the differences can be explained by changes in envelope structure and chemistry alone. However, we caution that this comparison is merely illustrative: as shown by the point colors, real exoplanets have a range of equilibrium temperatures. Additionally, the models presented in Fig.~\ref{fig:MR_diagram} are for a constant base temperature, and therefore do not correspond to a constant age, as low-mass planets should cool faster than their more massive counterparts. This leads to a rapidly increasing modeled radius at small masses, as noted in previous literature \citep[e.g.,][]{LF14, Rogers2023}, which complicates a direct comparison between the models and data that span a large range in ages as well as equilibrium temperatures. Our results thus indicate a need for comprehensive population synthesis that includes the effects of silicate condensation.

The sensitivity of our models at low masses suggests that they may predict divergent outcomes compared to previous models for small planets when they are at their youngest and largest, and when their interiors are at their hottest. We therefore expect that non-convective regions will significantly affect atmospheric escape, which is highly sensitive to planet radius. For example, changes in the mass-radius relation could affect the transition from core-powered to photoevaporative escape \citep{Misener2026}. Atmospheric escape is also dependent on atmosphere-interior interactions through compositional fractionation \citep{Cherubim2025} and impacts on the thermal profiles of the upper atmosphere \citep{MisenerSchulik2025}, both of which are affected by condensation deep in sub-Neptune envelopes. As these planets evolve and their interiors cool, the regions where convection is inhibited are no longer reached \citep{MisenerSchlichting2022}, decreasing the radius differences from these mechanisms, although the chemistry is altered in a similar fashion. Our results therefore show that planetary mass-radius assumptions are sensitive to the chemical makeup of the mantle and that a fully coupled model is vital for interpreting observed sub-Neptune bulk densities.

\subsection{Sensitivity to condensate speciation}\label{sec:speciation}
Our fully coupled models allow us to probe the sensitivity of predicted atmospheric abundances to the chemical constants, i.e. the Gibbs free energies, of species in the deep envelope, which are poorly constrained. As a case study, silicon oxide, SiO(s), is in the canonical FASTCHEM COND ensemble of potential condensates \citep{FastchemCond}, valid up to 6000~K \citep[see also][]{KitzmannStock2026}. However, it is not clear whether SiO(s) would really be expected at such depths, as its tabulated equilibrium constants were studied in the context of stellar winds, with pressures $\sim 10^{-4}$~bar and temperatures $\lesssim 1800$~K \citep{Gail2013SiO}. Other authors have also questioned this extrapolation to sub-Neptune interiors \citep{LeeWerlen2025}. To test our sensitivity to deep condensate speciation, we therefore compare the profiles we obtain including and removing SiO(s) from the chemical network.

\begin{figure*}
    \centering
    \includegraphics[width=\linewidth]{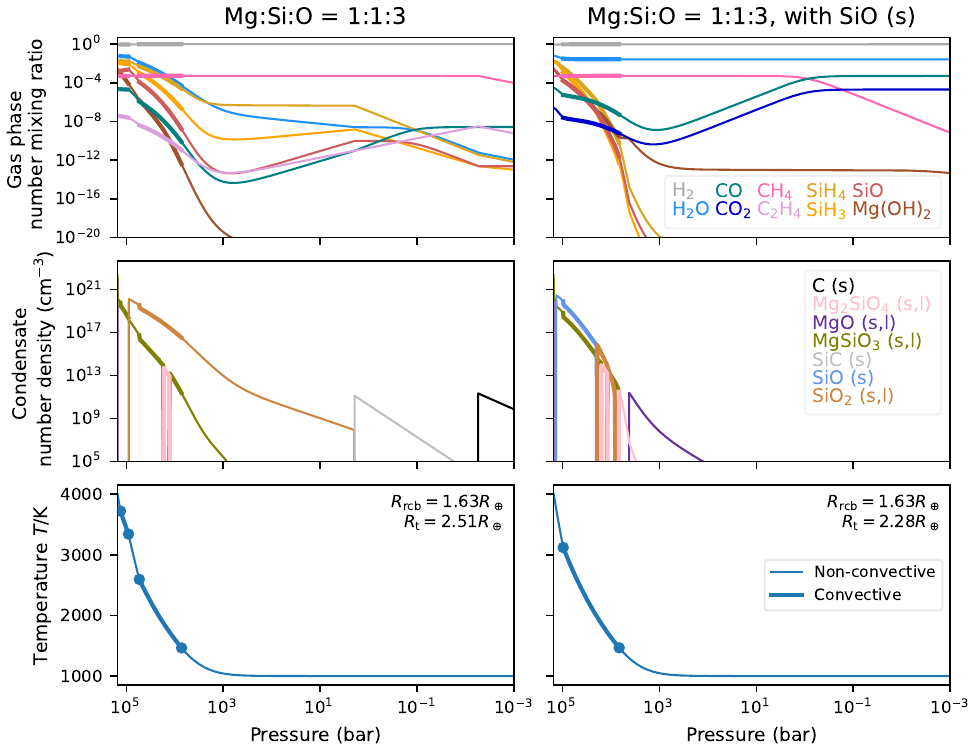}
    \caption{Comparison of atmospheric composition (top), condensates (middle), and temperature profile (bottom) for two models of a sub-Neptune with planet mass $M=4 M_\oplus$, equilibrium temperature $T_\mathrm{eq} = 1000$~K, base temperature $T_\mathrm{base}=4000$~K, and envelope hydrogen mass $f=0.03$, with an Mg:Si:O ratio of 1:1:3. The left column is identical to that in in Fig.~\ref{fig:MgSi_comp}, representing our fiducial model, while the right includes SiO(s). The inclusion of SiO(s) leads to markedly different internal structures and upper atmosphere compositions, underscoring the importance of relatively uncertain interior condensate species.
    }
    \label{fig:noSiOcondensate}
\end{figure*}

The inclusion of SiO(s) makes it the dominant silicate condensate over much of the deep envelope, in agreement with previous work \citep{Ito2025} (Figure~\ref{fig:noSiOcondensate}). In comparison to our canonical silicate condensates, SiO is much less efficient at removing oxygen from the gas phase, simply due to its lower O:Si ratio compared to e.g. MgSiO$_3$. Therefore, an envelope base composition (Mg:Si:O ratio of 1:1:3) that previously produced an oxygen-poor upper atmosphere without including SiO(s) now produces an oxygen-rich one, with a concomitant decrease in the transit radius. These results highlight the need to further understand the Gibbs energies and chemical properties of silicate condensates at these exotic temperatures and pressures through improved experiments and/or molecular dynamic modeling. Due to the questionable applicability of SiO (s) Gibbs energies to the high pressure, high temperature environments considered here, we continue to use the model without SiO (s) as our canonical suite of condensates throughout the rest of this work.

\section{Population trends}\label{sec:trends}
\subsection{Magma ocean composition}
\begin{figure*}
    \centering
    \begin{subfigure}{\columnwidth} 
       \includegraphics[width=\textwidth]{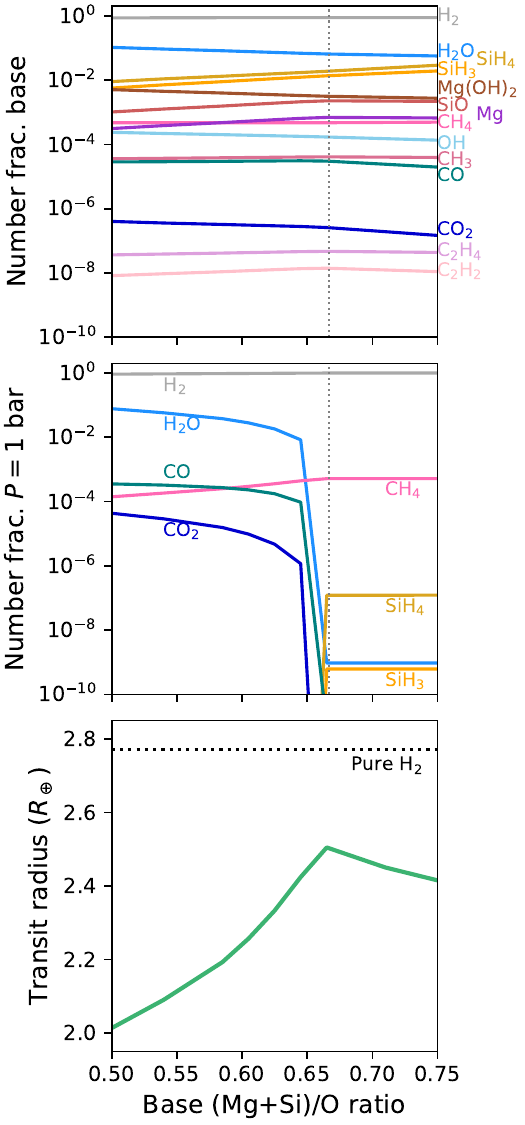}
    \end{subfigure}%
    \begin{subfigure}{\columnwidth} 
       \includegraphics[width=\textwidth]{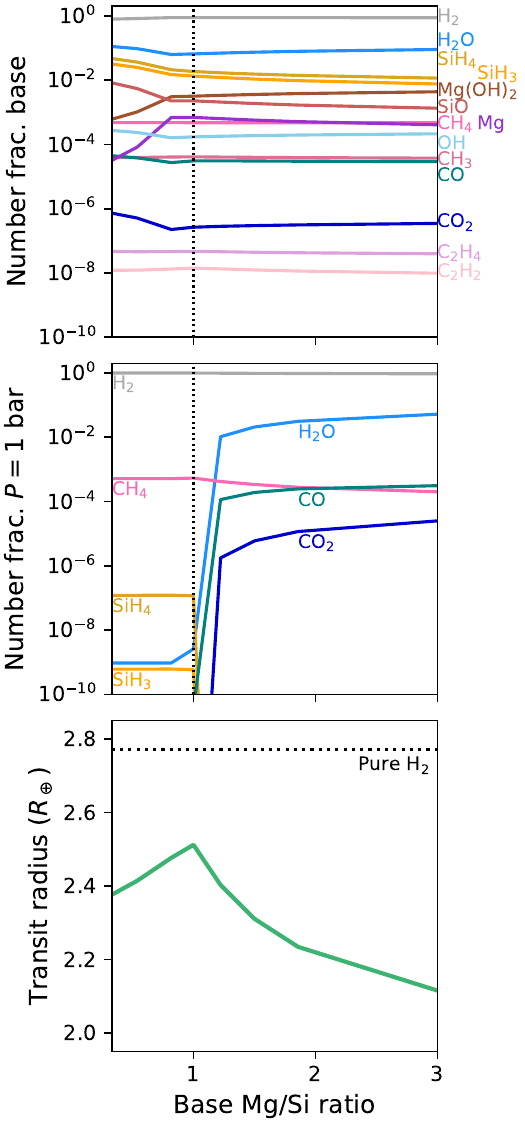}
    \end{subfigure}%
    \caption{Envelope base (top) and upper envelope (middle) gas abundances and transit radii (bottom) of a sub-Neptune with planet mass $M=4 M_\oplus$, equilibrium temperature $T_\mathrm{eq} = 1000$~K, base temperature $T_\mathrm{base}=4000$~K, and envelope hydrogen mass $f=0.03$. We vary two axes of interior composition: (Mg+Si)/O ratio (left), in which Mg/Si=1, and Mg/Si ratio (right), in which (Mg+Si)/O=2/3. Each vertical dotted lines show the value of each variable assumed in the other column. The horizontal dotted lines in the bottom row show the transit radius of an equivalent pure H$_2$ envelope. Along both compositional axes, gradual trends in the base composition lead to the upper envelope composition varying rapidly from an oxygen-poor endmember composition to an oxygen-rich one, with corresponding changes in the transit radius.
    }
    \label{fig:trends_comp}
\end{figure*}
To investigate the efficacy of atmospheric abundances as a probe of interior makeup, we explore two different magma ocean compositions: the (Mg+Si)/O and Mg/Si ratios. For bulk silicate Earth, these values are 0.61 and 1.2 respectively, but they are expected to be variable across the exoplanet population \citep[e.g.,][]{Putirka2021}. In both cases we observe a bifurcation in upper atmospheric composition: for relatively oxygen-rich and silicon-poor scenarios, i.e. low (Mg+Si)/O ratios and high Mg/Si ratios, the upper atmospheres have abundant volatiles with mixing ratios which trend slowly between more reducing and oxidizing compositions (Figure~\ref{fig:trends_comp}). These abundances are produced by silicate-limited condensation at depth, leading to oxygen-rich upper atmospheres, and thus correspond to the left column of Figure~\ref{fig:MgSi_comp}. For relatively oxygen-poor and silicon-rich compositions, conversely, we observe highly reducing atmospheres, with methane and silane predominating, corresponding to the right panel of Figure~\ref{fig:MgSi_comp}. Interestingly, the upper atmospheric composition does not vary smoothly between these two extremes, but rather transitions rapidly from oxidizing to reducing, here at (Mg+Si)/O$\sim 0.67$ and Mg/Si$\sim 1.3$, with only subtle variation within each regime. These results indicate that atmospheric composition can broadly probe the interior state, distinguishing between compositional endmembers.

Unlike the sharp discontinuities in the upper atmospheric abundances, the envelope base gas makeup varies smoothly with the changing magma ocean compositional ratios, trending weakly toward more reducing species with increasing (Mg+Si)/O and decreasing Mg/Si (Figure~\ref{fig:trends_comp}). In all cases, the dominant basal Si-bearing gas species is silane, SiH$_4$, and the dominant Mg-bearing gas species is magnesium hydroxide, Mg(OH)$_2$, similar to the results of non-ideal calculations \citep{GilmoreStixrude2026}. These findings thus demonstrate a tipping-point mechanism caused by multi-species condensation: upper atmospheric compositions are fundamentally dictated by the limiting elements in the condensate sequences, in this case either Si or O. When gas abundances are modified slightly, the ratio of condensates can change to accommodate it, resulting in similar total abundances, until the tipping point is reached and all of one element is exhausted.

The transit radii change non-monotonically as a function of basal magma composition (Figure~\ref{fig:trends_comp}), despite holding the total hydrogen mass constant. As shown previously, the radii change both with mean molecular weight and the thickness(es) of the non-convecting regions. The radii are largest close to the transition between condensates and smallest at each compositional endmember. For increasing oxygen and magnesium abundances, more oxygen remains in the upper atmosphere, increasing the mean molecular weight. Meanwhile, in the Si-rich and O-poor regime, with CH$_4$-SiH$_4$ dominated atmospheres, increasing the silicate abundance at the base leads to more extensive regions of condensation, larger super-adiabatic regimes, and thus smaller planets. These results demonstrate the complex interplay between silicate condensation and planetary radius.

\subsection{Equilibrium temperature}
\begin{figure*}
    \centering
    \begin{subfigure}{\columnwidth} 
       \includegraphics[width=\textwidth]{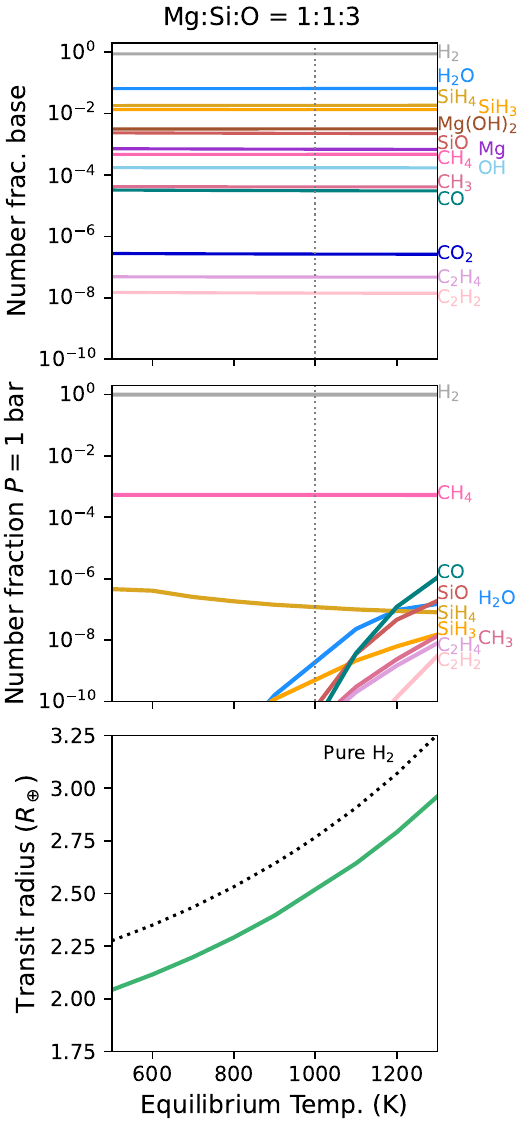}
    \end{subfigure}%
    \begin{subfigure}{\columnwidth} 
       \includegraphics[width=\textwidth]{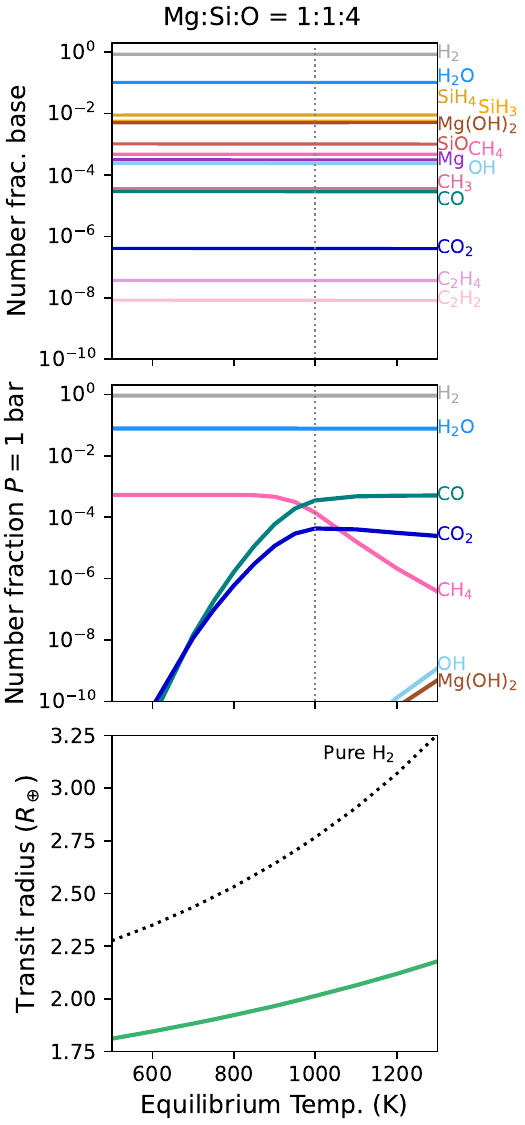}
    \end{subfigure}
    \caption{Envelope base (top) and upper envelope (middle) gas abundances and transit radii (bottom) as functions of equilibrium temperature with fixed mass $M=4 M_\oplus$, base temperature $T_\mathrm{base}=4000$~K, envelope hydrogen mass $f=0.03$, and magma Mg:Si:O=1:1:3 (left) and 1:1:4 (right). The vertical dotted lines in the top two rows indicate the nominal equilibrium temperature considered above, 1000~K. The dotted curves in the bottom row are for pure H envelopes. In both cases, the upper atmosphere transitions to a more oxygen-rich composition with increasing equilibrium temperature and fixed magma composition, while the base atmospheric composition, determined only by local pressure, temperature, and composition, is independent of the equilibrium temperature.
    }
    \label{fig:trends_T}
\end{figure*}

Changes in equilibrium temperature do not impact the envelope base gas composition, as it is determined solely by pressure, base temperature, and magma composition, all of which are unchanged by changing equilibrium temperature (Figure~\ref{fig:trends_T}). In contrast, upper envelope abundances vary with equilibrium temperature depending on envelope base conditions. For the O-rich case, 
we find the well-known CH$_4$-CO transition at $\sim 900$~K \citep[e.g.,][]{PrinnBarshay1977, Moses2013, FortneyVisscher2020}, showing that such trends are still apparent in this coupled modeling framework. The upper atmospheric composition changes come despite the base compositions being fully independent of equilibrium temperature. 
For the O-poor case, we see an increase in the abundances of species beside CH$_4$ and SiH$_4$, including CO and SiO, as temperature increases. Therefore, we predict that hotter sub-Neptunes may provide an opportunity to measure more species-rich atmospheric spectra that can help better pin down the deep envelope composition. In cases with non-standard condensates, such as SiO(s), we find additional diverse speciation, including atomic Mg. These low-oxygen envelopes also favor production of complex hydrocarbons, which could lead to soot formation \citep{YangKempton2026} and active photochemistry \citep{Morley2015}.

The transit radii display expected behavior as equilibrium temperature is varied: in the oxygen-poor case, our transit radii are offset to lower values from the pure H case due to the presence of super-adiabatic regions, but scale similarly with temperature due to their common scale height dependence on $T_\mathrm{eq}$. Meanwhile in the oxygen-rich case, the higher mean molecular weight means that transit radii increase more slowly with temperature than in the pure H endmember.

As we do not include volatile condensates in this work, we limit ourselves to equilibrium temperatures $\geq 500$~K to stay above the condensation temperature of the major expected species, notably water. Water condensation would decrease the oxygen abundances in cool sub-Neptunes, as has been invoked to explain the water-poor composition of K2-18~b, for example \citep[e.g.,][]{Madhusudhan2023}.

\subsection{Atmospheric mass}
\begin{figure*}
    \centering
    \begin{subfigure}{\columnwidth} 
       \includegraphics[width=\textwidth]{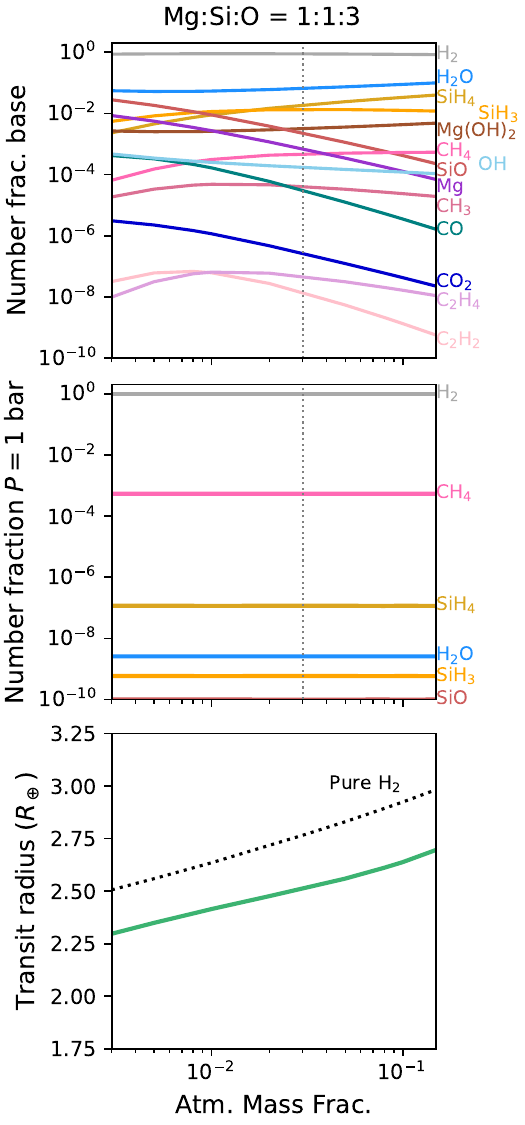}
    \end{subfigure}%
    \begin{subfigure}{\columnwidth} 
       \includegraphics[width=\textwidth]{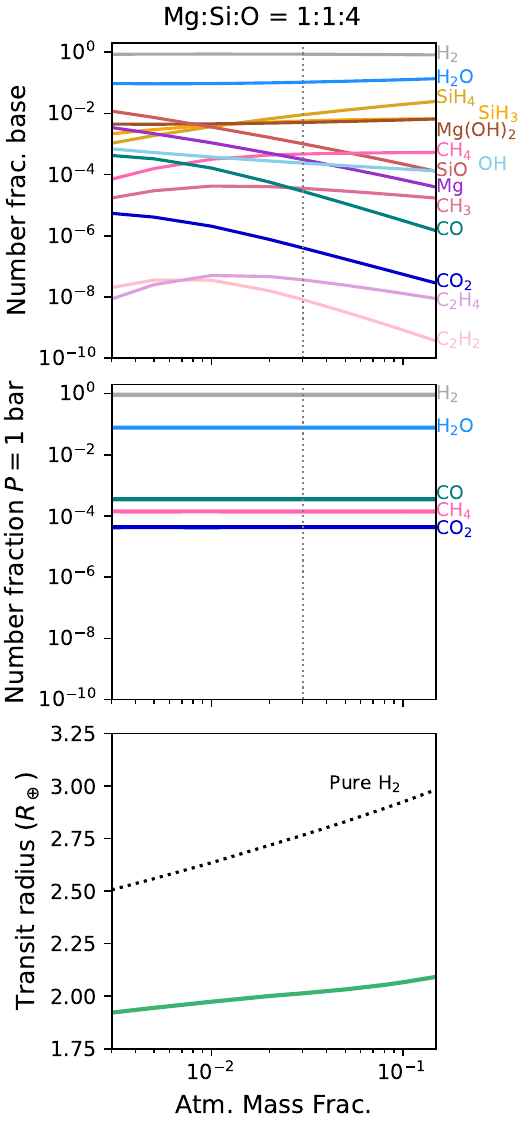}
    \end{subfigure}
    \caption{Envelope base (top) and upper envelope (middle) gas abundances and transit radii (bottom) as functions of atmospheric mass fraction  with fixed mass $M=4 M_\oplus$, base temperature $T_\mathrm{base}=4000$~K, equilibrium temperature $T_\mathrm{eq}=1000$~K, and magma Mg:Si:O=1:1:3 (left) and 1:1:4 (right). The vertical dotted lines in the top two rows indicate the nominal atmospheric hydrogen mass we have considered thus far, $f=0.03$. The dotted curves in the bottom row are for pure H envelopes. The upper atmospheric composition is virtually independent of the atmospheric mass fraction, while the base composition is highly variable, becoming more reducing with increasing hydrogen fraction.
    }
    \label{fig:trends_f}
\end{figure*}

\begin{figure*}
    \centering
    \begin{subfigure}{0.5\textwidth} 
       \includegraphics[width=\textwidth]{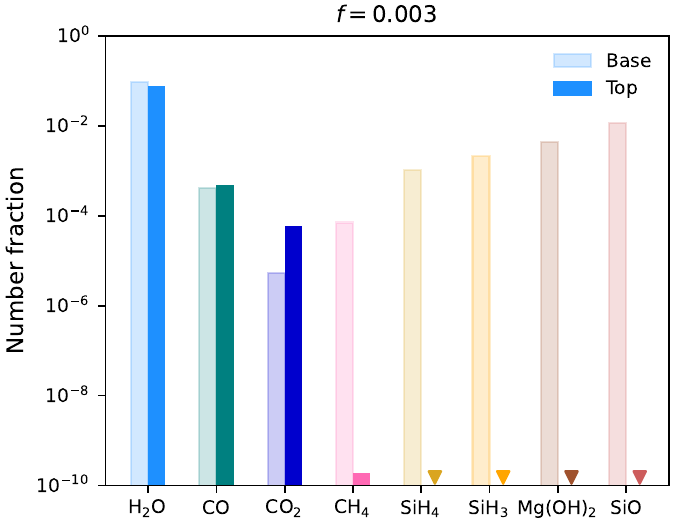}
    \end{subfigure}%
    \begin{subfigure}{0.5\textwidth} 
       \includegraphics[width=\textwidth]{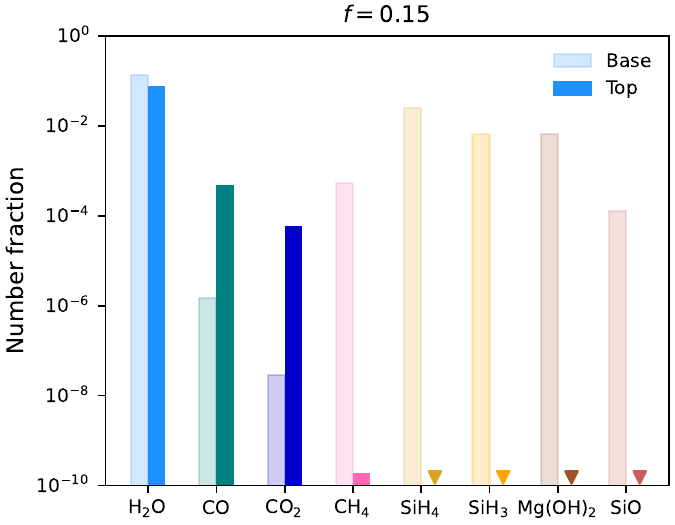}
    \end{subfigure}%
    \caption{Comparison of the atmospheric gas composition at the base (lightly shaded bars) and top (heavily shaded bars) of the atmosphere for two models with fixed mass $M=4 M_\oplus$, base temperature $T_\mathrm{base}=4000$~K, magma composition (Mg:Si:O=1:1:4), and equilibrium temperature $T_\mathrm{eq}=1000$~K but different atmospheric mass fractions, $f=0.003$ (left) and $f=0.15$ (right). Downward triangles denote abundances below the lower bound of the plot. The low envelope mass has a more oxidized composition at the base, due to the lower hydrogen pressures. However, condensation removes this oxygen more efficiently, leading to virtually identical compositions in the visible layers.
    }
    \label{fig:bottom_top}
\end{figure*}

Atmospheric mass variations affect abundances in the opposite way as equilibrium temperature: the envelope base composition varies dramatically with $f$, becoming significantly more reducing with increased atmospheric mass fraction (hydrogen gas pressure), but the upper atmosphere makeup is not impacted (Figure \ref{fig:trends_f}). While these results appear to show insensitivity of the upper atmosphere to bulk properties of the planet, they actually point to a novel intrinsic damping effect from deep condensation.
This is demonstrated more clearly in Figure~\ref{fig:bottom_top}, where we compare the composition of the most abundant oxygen-, carbon-, and silicon-bearing species at the base and top of the hydrogen envelopes for two extreme values of atmospheric hydrogen mass $f$, 0.003 and 0.15. The compositions at observable (low) pressures are essentially identical between the two cases, matching those shown in Fig.~\ref{fig:trends_f}. In contrast, the compositions at the base of the envelope differ by orders of magnitude. Due to the much lower hydrogen pressure, the $f=0.003$ case is much more oxidizing, with abundant gaseous CO, SiO, and Mg(OH)$_2$. In the $f=0.15$ case, the total equilibrium gaseous oxygen number density is actually slightly higher at the base of the envelope, but the much higher hydrogen pressure shifts the equilibrium toward SiH$_4$ and CH$_4$, i.e. a much lower oxygen fugacity. These different chemical equilibria result in different condensate sequences in the deep envelope. In the $f=0.003$ case, the deep condensate is mostly MgSiO$_3$, which evenly depletes Mg and Si along with O. In the $f=0.15$ case, in addition to MgSiO$_3$ condensation there is a much larger region of SiO$_2$ condensation, which occurs at temperatures between 2000 and 3500~K. This SiO$_2$ condensation, favored due to the lower oxygen fugacity and higher SiH$_4$ abundance, rapidly depletes the SiH$_4$ and oxygen levels until $\sim 1800$~K, at which point the equilibrium condensate switches to Mg$_2$SiO$_4$ and then MgO at similar temperatures to the $f=0.003$ case. In other words, rather than amplifying any initial compositional differences due to different envelope base pressure, the condensation sequence acts as a negative feedback: the condensation of SiO$_2$ increases to absorb the excess Si and O, taking both compositions to the same equilibrium in the upper envelope.

In contrast to Figure~\ref{fig:trends_comp}, in which small changes in base composition lead to drastic changes in upper atmospheric composition, Figures~\ref{fig:trends_T} and \ref{fig:trends_f} show how trends in upper atmospheric composition can mask basal compositions, either by varying with unchanging basal composition or by not varying despite major changes in the base composition. Both of these cases highlight the importance of fully modeling these envelopes from interior to atmosphere, rather than modeling the atmosphere as a `box' with constant composition or deriving their envelope radii independently.

\section{Observable consequences} \label{sec:observability}

\begin{figure*}
    \centering
    \includegraphics[width=1\linewidth]{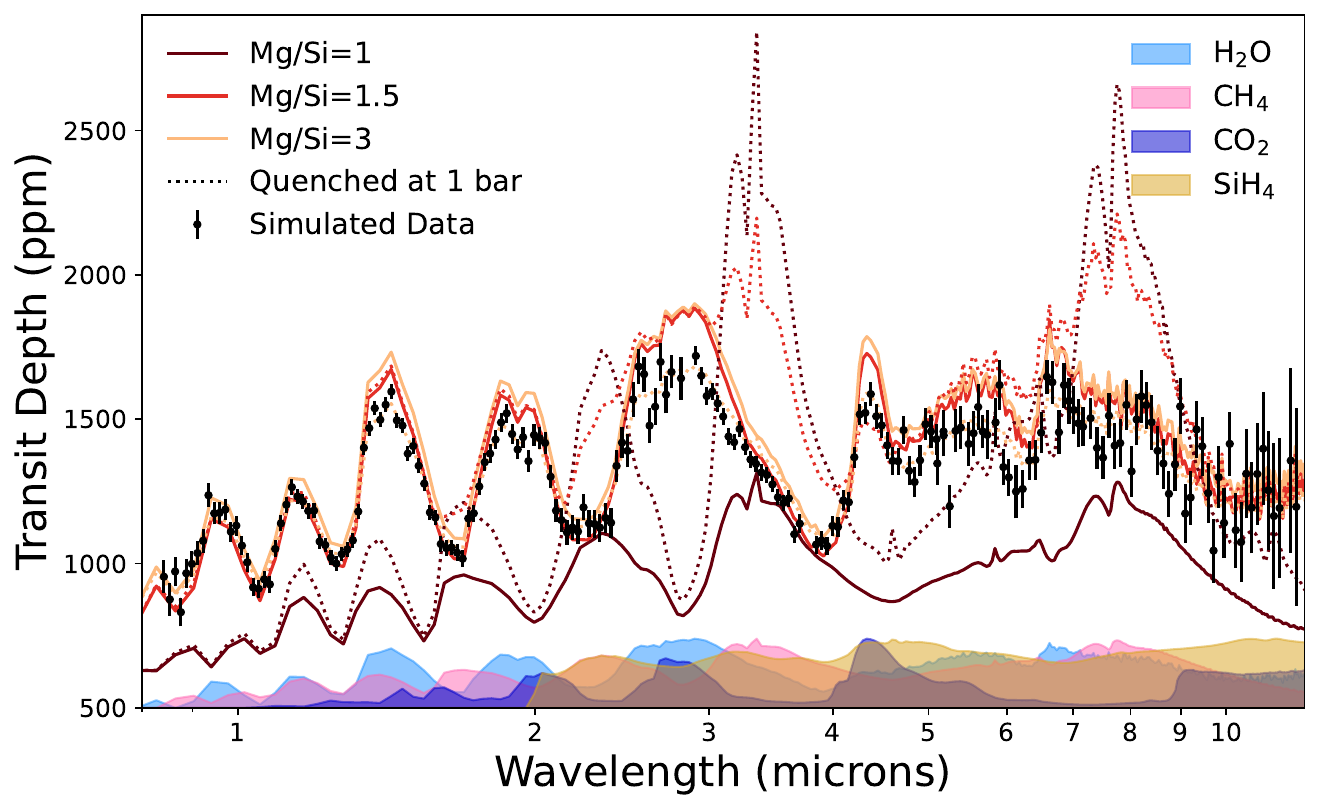}
    \caption{Model spectra computed assuming three different magma Mg/Si ratios (colors) in equilibrium (solid lines) and quenched at 1 bar (dotted lines) of hot sub-Neptune TOI-544~b. Black points show simulated observations of the quenched Mg/Si=3 case (light orange dotted model) using \texttt{PandExo}, assuming one transit in each of \textit{JWST} NIRISS/SOSS, NIRSpec/G395H, and MIRI all binned to $R \sim 40$. Shaded curves along the bottom show the opacities of the most relevant species. We observe a wide variety of spectral features, including H$_2$O, CO$_2$, CH$_4$, and SiH$_4$, that are readily distinguishable by \textit{JWST}, indicating that the basal compositions are inferrable from transmission observations. We also find that our model spectra are highly sensitive to the quench pressure in the upper atmosphere.
    }
    \label{fig:Spectra}
\end{figure*}

In order to demonstrate the observable consequences of the condensate sequences we consider in this work, we generate model spectra using the \texttt{Rocky Raccoon} abundance profiles for three magma Mg/Si ratios with (Mg+Si)/O=2/3 (Fig.~\ref{fig:Spectra}). These ratios produce profiles similar to those shown in Fig.~\ref{fig:trends_comp}, but here we assume the specific system parameters of the hot sub-Neptune TOI-544~b, a $2.89~M_\oplus$, $2.02 R_\oplus$ planet with an equilibrium temperature of 1115~K, assuming zero albedo, orbiting a K dwarf \citep{Osborne2024}. We first generate a series of equilibrium chemistry models for three representative thermal and chemical profiles using \texttt{Rocky Raccoon}. Using these thermochemical profiles, we generate model transmission spectra using \texttt{petitRADTRANS} \citep{pRT}. Additionally, we generate a series of models in the same way assuming the composition of the photosphere is wholly set at 1~bar, a conservative approximation of disequilibrium chemistry which we justify further in Sec.~\ref{sec:discussion}.
We use the default opacities from \texttt{petitRADTRANS} with the exception of SiH$_4$, which we add manually via the pre-computed \texttt{petitRADTRANS} format from ExoMol \citep{Owens17ExoMolSiH4}. Measured SiH$_4$ opacities only exist for wavelengths $> 2$~$\mu$m \citep[Fig.~\ref{fig:Spectra}; see also][]{Ito2025}. Therefore we do not include any SiH$_4$ absorption shortward of 2~$\mu$m, which we speculate would be qualitatively similar to isostructural methane but shifted to longer wavelengths. We then simulate the observations of a representative model, the Mg/Si=3 case, using \texttt{PandExo} \citep{pandexo} assuming 1 transit of TOI-544~b per instrument for each of NIRISS/SOSS using SUBSTRIP96, NIRSpec/G395H, and MIRI in the slitless LRS model. We do not include any aerosols, which have been proposed to be favored for sub-Neptunes with high C/O and high equilibrium temperatures \citep{YangKempton2026}.

Our model spectra show that the different interior compositions are clearly distinguishable at the precision of the simulated data (Fig.~\ref{fig:Spectra}). In particular, the low Mg/Si ratio case (Mg/Si=1) produces a CH$_4$ and SiH$_4$ dominated envelope, in line with the trend seen in Fig.~\ref{fig:trends_comp}. However, SiH$_4$ features are largely absent in the equilibrium spectra due to SiC rainout in the equilibrium model (Fig.~\ref{fig:MgSi_comp}). With virtually no oxygen-bearing species present, we thus predict spectra made up primarily of large CH$_4$ features, with low metallicities ($\sim 0.35 \times$ solar) and high inferred C/O ratios. The lack of molecular absorption aside from methane in the equilibrium Mg/Si=1 model means that transmission observations would probe deeper into the envelopes, leading to decreased transit depths.
Consideration of quenching produces a small SiH$_4$ feature in the spectrum at $\sim 4.5$~$\mu$m, suggesting that the prominence of this `silane finger' depends sensitively on the quench pressure assumed. This feature has been observed by \textit{JWST} in the cold brown dwarf W1534 \citep{FahertyMeisner2025}, where the SiH$_4$ abundance was constrained to $\sim$20~ppb, lower than that predicted here. Quenching also greatly increases the prominence of the methane feature (Fig.~\ref{fig:Spectra}, dotted lines). These atmospheres may be sensitive to photochemical effects, although the photochemical behavior of silane is highly uncertain, as we discuss further in Section~\ref{sec:disequil}.

The higher Mg/Si cases produce oxygen-rich envelopes containing abundant H$_2$O and CO$_2$, with the larger CO$_2$ features in the Mg/Si=3 case corresponding to the increasing CO$_2$ abundance with Mg/Si ratio shown in Fig.~\ref{fig:trends_comp}. These envelopes have super-solar metallicities ($\sim 25 \times$ solar) with sub-solar C/O ratios. These high abundances overcome the smaller scale heights compared to the low Mg/Si model, producing similar feature sizes as in the methane-rich case detectable with \textit{JWST} precision. As seen on the right side of Fig.~\ref{fig:MgSi_comp}, the equilibrium between CH$_4$ and CO/CO$_2$ switches close to 1~bar, explaining why we find prominent methane features in the quenched Mg/Si=1.5 case but not in the equilibrium case \citep[e.g.,][]{FortneyVisscher2020}.

Hot, small sub-Neptunes like TOI-544~b are most optimal for observing silane due to their large scale heights, but as demonstrated in Sec.~\ref{sec:trends}, we expect observable outcomes of atmosphere-interior interactions across the sub-Neptune population. Thus far, sub-Neptune population trends have been elusive, with numerous featureless spectra \citep[e.g.,][]{GaoPiette2023, Alderson2024, Wallack2024}. The few planets with detectable features have been highly diverse, with inferred mean molecular weights ranging from solar \citep{Davenport2025} to notably higher \citep{BennekeRoy2024, PiauletGhorayeb2024}, similar to the range in our models. However, few planets yet have well-constrained C/O ratios. As more observations are taken and trends emerge, the predictions of these models can be tested empirically.

\section{Discussion} \label{sec:discussion}
Our results provide motivation for understanding the major uncertainties in the physical and chemical properties of sub-Neptune interiors. We discuss a few of these uncertainties, including assumptions about ingassing, equilibrium chemistry, and atmospheric structure, below.

\subsection{Deep chemical processing in sub-Neptunes}
We have explored a range of chemical speciations in the silicate interior, but we remain agnostic to the formation processes which lead to each possible composition. While our fiducial models represent Mg, Si, and O values similar to bulk silicate Earth \citep{Palme2003}, a large range is possible in the exoplanet population, due both to intrinsic spread in stellar abundances \citep[e.g.,][]{Putirka2021}, formation processes such as the accretion of oxygen-rich ices, and planetary chemical equilibrium. Additionally, while we maintained a constant, solar carbon abundance in this work, exoplanetary bulk carbon abundances are likely strong tracers of formation beyond the carbon-ice lines \citep[e.g][]{YangHu2024, Steinmeyer2026}. Coupling a full interior chemical equilibrium model \citep[e.g][]{SY22, BowerThompson2025, GrimmSteinmeyer2026} to the bottom of the models presented here would more thoroughly link bulk composition to atmospheric observables, equivalent to specifying a location in e.g. the Mg/Si and oxygen abundance plane, and is the subject of planned future work. 

We have always assumed silicate magma lies at the base of the envelopes. However, improved \textit{ab initio} and chemical modeling has suggested significant mixing of hydrogen and carbon into melts within sub-Neptunes \citep{BowerThompson2025, RogersDorn2025SEwater, Ito2025, Werlen2025C-O} and even full miscibility of hydrogen and silicates in a super-critical phase \citep{YoungStixrude2024, RogersYoung2025miscible, GilmoreStixrude2026} close to the temperatures and pressures we consider. Miscible regions are highly complementary to the physics we treat here, which focus on the structure and chemistry of the hydrogen-dominated layers still expected above them. 

The effects of miscibility are impossible to consider fully without accounting for non-ideality of the gas and melt species. The assumption of ideal behavior in this work is a simplification, as recent work has demonstrated that the H/He-Si-O mixtures expected in sub-Neptunes begin deviating from ideality at $\sim$~kbar pressures \citep[e.g.,][]{BowerThompson2025}. Non-ideal effects can change the energetics of the system in myriad ways, including altering ingassing efficiencies \citep{BowerThompson2025, WerlenYoung2026}, perturbing the temperature-pressure profile \citep{RogersYoung2025miscible}, and modifying gas chemical equilibrium \citep{HakimBower2026}. Specifically, \citet{HakimBower2026} found that including real gas effects, such as a novel SiH$_4$ equation of state, strongly favors the production of SiH$_4$ deep in sub-Neptune envelopes while disfavoring oxygen-bearing species, leading to SiH$_4$-CH$_4$ dominated envelopes similar to and in some cases more extreme than those we consider here in the reducing cases. This could promote reduced silicon-bearing condensates like Si, but the effects of magnesium bearing species were not considered. However, incorporating non-ideal corrections to the activities of the melt species in addition to the gases dampens these effects, promoting oxygen content in the envelope to levels relatively close to ideal approximations \citep{WerlenYoung2026}. While FASTCHEM COND assumes ideality, approaches that consider non-ideality of a large number of gas and melt species could lead to interesting insights. However, though the exact chemical-thermal profiles may change under non-ideal conditions, we expect that the physical principles we demonstrate here remain true: consideration of multiple condensates and self-consistent chemistry and structure fundamentally alters the interiors and upper atmospheres of sub-Neptunes compared to models with limited condensate speciation.

Ingassing is thought to have a variety of competing effects on sub-Neptunes. One possibility is that the melt's activity could decrease, leading to sub-saturation above the magma ocean surface and thus a deep, high mean molecular weight convective region in the envelope that persists until an environment favoring saturation and condensation is reached. In addition, efficient water ingassing could decrease the water abundance in the atmosphere \citep{BowerThompson2025, Werlen2025waterworlds}, further favoring SiH$_4$ formation \citep{Ito2025, HakimBower2026}. Ingassing has also been proposed to have major effects on the partitioning of gas-phase elements we do not consider in this work, including sulfur and nitrogen \citep[e.g.,][]{ShorttleJordan2024, WerlenBurn2026}. Ingassing likely also changes the radii of the planets below the envelope-interior boundary, and therefore alter our conclusions on their mass-radius relationships. Increases in planet radii due to ingassing of hydrogen are expected to be on the order of 10\% \citep{RogersYoung2025miscible}, with further changes of similar magnitude due to the miscibility of iron, hydrogen, and silicates \citep{YoungWerlen2025}. These changes to radii are comparable to those expected from the chemical gradients we consider here, but ingassing also draws down hydrogen from the envelope. Therefore, we still expect planets with the same bulk hydrogen to be smaller than a simpler model that considers uncoupled layers.

\subsection{Which clouds actually form?}\label{sec:disequil}
As demonstrated above, the specifics of which condensates form throughout sub-Neptune envelopes play a major role in determining the planet radii and observable composition, but are highly uncertain. In this work we were forced to extrapolate the Gibbs energies -- and thus the equilibrium speciation -- far beyond their experimentally verified values. The case study of SiO(s) (see Figure~\ref{fig:noSiOcondensate}) exemplifies this, but the argument holds for any of the species we consider here. We therefore hope that this study spurs further refinement of the relevant Gibbs energies at the relevant temperatures and pressures. In addition, equilibrium condensation is in itself an assumption: nucleation barriers can inhibit condensate formation in some environments \citep[e.g.,][]{gao2020, LeeWerlen2025}. However, in the dense, hot environments at the great depths that we focus on here, we do not expect kinetic and nucleation barriers to be significant. Meanwhile, condensates at depth could also be lofted instead of raining out, altering observables \citep{LeeWerlen2025} and thermal profiles \citep{MukherjeeNixon2026}.

Our results show that rock vapor clouds at depth are not the only condensates affecting the observable spectra. In some reducing cases we find that SiC condensation can draw down SiH$_4$ in the atmosphere at pressures $\lesssim 1$~bar (see Figure~\ref{fig:MgSi_comp}). SiC condensation was mentioned in \citet{Ito2025} as a possible mechanism for depleting atmospheric carbon abundances in silane dominated sub-Neptunes, but here we model envelopes with sufficient carbon to deplete silicon instead. Silicon hydride gases are also thought to be important in the formation of SiC interstellar grains in AGB stars \citep{Tajuelo-Castilla2026}, although those calculations are performed at much lower hydrogen pressures than we consider here. 

Going beyond equilibrium chemistry for silicon hydrides is difficult, as the kinetics of the relevant reactions, and indeed of SiH$_4$ in sub-Neptune atmospheres, are not well understood. Measurements of SiH$_4$ decomposition in an argon background gas indicate a rate constant of $k \sim 10^{6}$~cm$^3$\ mol$^{-1}$\ s$^{-1}$ at 1000~K \citep{Petersen2003silanekinetics}. Eddy diffusion coefficients in sub-Neptunes are highly uncertain, but for those expected in sub-Neptune atmospheres \citep{Charnay2015}, these rate constants translate to a quench pressure of 3~bar for a weakly-mixing case ($K_{zz}=10^6$~cm$^2$\ s$^{-1}$) and 150~bar for a strongly mixing case ($K_{zz}=10^9$~cm$^2$\ s$^{-1}$), justifying our conservative quench pressure in Sec.~\ref{sec:observability}. However, these and other literature measurements of silane properties are usually performed in neutral background gases or in environments relevant to the semiconductor industry \citep[e.g.,][]{Petersen2003silanekinetics, AccollaSantoro2021, Szafarska2023silanechemistry}, not the dense, highly reducing hydrogen envelopes we consider here. Additionally, the condensation of silane into SiC may follow a chain of reactions, where decomposition may not be the rate-limiting step, as in the conversion of SiH$_4$(g) to SiO(g) \citep{FegleyLodders1994, FahertyMeisner2025}, implying yet deeper quenching. Silane abundances in oxygen-rich scenarios may also be affected by the formation of silicon hydroxides \citep[e.g.,][]{Fegley2016, Bauschlicher2023, Szafarska2023silanechemistry}, which are not present in the FASTCHEM COND chemical network and so are not treated in this work. Nonetheless, in oxygen-rich scenarios the silicon rains out quickly, indicating that silicon hydroxides are likely unimportant observationally. 

The abundances of the main carbon- and oxygen-bearing species we consider are also likely susceptible to disequilibrium chemistry in the upper atmosphere. As shown in Figure~\ref{fig:Spectra}, for some temperature profiles, chemical quenching favors disequilibrium methane over CO and CO$_2$, typically at pressures between $10^{-2}$ and $10^2$~bar depending on mixing strength \citep[e.g.,][]{FortneyVisscher2020}. The oxygen-poor, methane-rich envelopes we find could also promote formation of photolytic \citep{Morley2015, gao2020, GaoPiette2023} or deep carbon chemistry-based \citep{YangKempton2026} aerosols, potentially muting spectral signatures. Such reactions would also concomitantly deplete methane in favor of higher-order hydrocarbons; it is plausible that similar reactions could occur for silane as well. Therefore, further work on the detailed kinetics and chemistry of the methane-silane envelopes predicted here is needed to fully assess their observability.

\subsection{Atmospheric structure and evolution: open questions}
Our work posits the widespread existence of non-convective regions for any planet that reaches 2500~K in its hydrogen-dominated envelope with underlying silicates, which includes most sub-Neptunes \citep{MS21}. This stems from our `equilibrium structure' assumption, i.e. that the hydrodynamics of sub-Neptune envelopes do not affect our results. However, the 3D stability of non-convective structures at the base of sub-Neptune envelopes is not well constrained. General circulation models investigating similar non-convective regions due to water condensation have shown them to be stable against dynamical effects \citep{Leconte2024, Habib2024}, but these non-convective regions occur at much lower temperatures and pressures than we consider. Possible `bursty' convection with periodic storms that pierce the non-convective region, as inferred for Saturn \citep{LiIngersoll2015} and Uranus and Neptune \citep{Clement2024}, would provide an intriguing way of releasing higher mean molecular weight species into the detectable atmosphere.

Our super-adiabatic gradients are sensitive to the heat flux being transferred through the non-convective region, which we assume to be equal to the flux out of the planet. However, at early stages in a planet's evolution, some of the planetary cooling goes into the envelope's contraction rather than interior cooling, reducing $L$ in the deep regions and, consequently, increasing the lapse rate per Eq.~\ref{eq:radtemp}. The internal heat flux at depth may also increase, due to the deposition of tidal heating \citep[e.g.,][]{Millholland2019} and the condensation of silicates itself, which releases both latent heat and gravitational potential energy \citep[e.g.,][]{VazanOrmel2024, RogersYoung2025miscible, TejadaArevaloGupta2026}. The actual value of $L$ at depth cannot be adequately analyzed without a time-evolving model, which we leave for future work.

A time-evolving model should additionally consider the interactions between atmospheric escape and our structure results. Sub-Neptune evolution and population synthesis models used to draw conclusions about, e.g., atmospheric escape typically assume convective hydrogen envelopes \citep[e.g.,][]{OW17, GS19, RogersGupta2021}. Smaller-radius envelopes are deeper in their planets' gravitational potentials and are therefore less prone to atmospheric escape. Changes in the envelope mass-radius relation can thus alter evolutionary predictions, such as the timing of different phases of atmospheric escape \citep{Misener2026}. These changes in radius ought to be most drastic when the planets are young and hot at depth, during their early phases of escape, which sets the stage for the rest of their evolution \citep{Rogers2024}. Furthermore, we neglect the radiative feedback of different chemical speciation in the upper atmosphere on the temperature profiles, which requires consideration of wavelength-dependent opacities and could significantly impact atmospheric loss rates \citep{MisenerSchulik2025}. Finally, as escape occurs, the envelope base pressure decreases, potentially allowing for the outgassing of hydrogen, buffering against envelope mass loss \citep[e.g.,][]{CS18, Steinmeyer2026}, as well as other species. The escape itself may also change the composition of the bulk atmosphere, due both to fractionation in the wind \citep{Cherubim2025, Valatsou2026} and to the composition in the upper atmosphere, from which the wind is launched, not being representative of the entire envelope. Changing compositions over time due to the coupling of escape and outgassing has major implications for planetary evolution and thus the super-Earth and sub-Neptune population, making it a highly interesting topic of future study. To summarize, our findings here show rich potential for integration into general planetary evolution models.

\section{Conclusion} \label{sec:conclusion}
In this work, we present a novel coupled atmospheric structure-chemical model for sub-Neptune exoplanets: \texttt{Rocky Raccoon}. We incorporate Mg, Si, C, O, and H-bearing species, a significant advance over previous coupled models. Our major conclusions are as follows:
\begin{itemize}
    \item The condensation sequences of magnesium-silicates within sub-Neptune envelopes are determined by a combination of basal magma composition, envelope hydrogen mass, and envelope base temperature. These condensation sequences drastically alter the upper atmospheric elemental abundances and can lead to the formation of multiple non-convective regions in the deep envelope.
    \item Condensation bifurcates the chemical abundances of the observable atmosphere of sub-Neptunes into two endmembers: if oxygen efficiently condenses out, the atmospheres are low mean molecular weight ($\mu \sim 2$~amu) and rich in methane and silane. Conversely, oxygen-rich interiors condense out magnesium and silicon and leave water and carbon monoxide-rich atmospheres with $\mu \sim 4$~amu. These atmospheric compositions are readily distinguishable with \textit{JWST} observations.
    \item The locations and widths of the non-convective regions vary with condensation sequence, affecting interior transport and envelope structure and ultimately altering the envelope mass-radius relation. As these non-convective regions are typically super-adiabatic, they make sub-Neptunes at least 10\% smaller than a fully convective model, suggesting that many sub-Neptunes contain more massive envelopes than previously predicted. The envelope structures computed here deviate markedly from fully adiabatic, constant molecular weight structures.
    \item We find an abrupt transition in sub-Neptune envelope composition and molecular weight from oxygen-rich to oxygen-poor at magma composition ratios of (Mg + Si)/O $\gtrsim 0.67$ and Mg/Si $\lesssim 1.05$ for our model choices. These transitions are due to changes in the condensate sequence, which reach tipping points despite smooth variations in the basal gas speciation.
    \item Compositional transitions in the upper atmosphere are highly sensitive to which condensates are stable at depth, with changes in the condensate speciation leading to wholesale differences in the observable atmospheric composition. However, we are typically forced to extrapolate the condensate equilibrium constants to the pressures (from $\sim 10^4$ to $10^5$~bar) and temperatures (from $\sim 2000$\ to $>4000$~K) relevant to sub-Neptune interiors. Laboratory experiments and molecular dynamics modeling are needed to refine these equilibria and determine which species make it to the observable layers of sub-Neptune envelopes. Observations of unusual species predicted here like atomic Mg may also signal unexpected condensate sequences.
\end{itemize}
Our results have major implications for our understanding of sub-Neptune formation and evolution. The bifurcation in atmospheric composition constrains existing and future observations to specific regions of interior abundance parameter space. Meanwhile, the larger envelope masses/smaller-radius envelopes we find imply that sub-Neptunes are less susceptible to atmospheric escape than previously assumed, perhaps pointing to efficient ingassing. Though we limited ourselves to a five element system, our new framework is easily extensible to more complete elemental compositions. Future work incorporating atmospheric evolution, deep chemical mixing, and more realistic silicate interior compositions into our envelope structure-chemistry models and comparing their findings to observed exoplanet spectra will be key to significantly furthering our understanding of the small exoplanet population.

\begin{acknowledgments}
We thank the reviewer for thoughtful comments that improved the manuscript. This AEThER research is funded in part by the Alfred P. Sloan Foundation under grant G-2025-25284. This research has made use of the NASA Exoplanet Archive \citep{PSCompPars}, which is operated by the California Institute of Technology, under contract with the National Aeronautics and Space Administration under the Exoplanet Exploration Program. This work made use of \texttt{OverCite} \citep{Shariat2026}, an in-editor citation tool for \LaTeX.

\end{acknowledgments}

\software{NUMPY \citep{numpy},
          MATPLOTLIB \citep{Matplotlib},
          SCIPY \citep{scipy},
          FASTCHEM COND \citep{FastchemOriginal, Fastchem2, FastchemCond}}

\bibliography{si_vapor}{}
\bibliographystyle{aasjournalv7}

\appendix
\section{Species included in chemical equilibrium} \label{sec:appendix_tables}
\begin{longtable}{ccc}
\textbf{Gas Species} & \textbf{Species Name} & \textbf{Source} \\
\endfirsthead
\hline
C$_1$H$_1$          & Methylidyne              & JANAF, \citet{NIST} \\
C$_1$H$_1$O$_1$     & Formyl                   & JANAF, \citet{NIST} \\
C$_1$H$_2$          & Methylene                & JANAF, \citet{NIST} \\
C$_1$H$_2$O$_1$     & Formaldehyde             & JANAF, \citet{NIST} \\
C$_1$H$_3$          & Methyl                   & JANAF, \citet{NIST} \\
C$_1$H$_4$          & Methane                  & JANAF, \citet{NIST} \\
C$_1$H$_4$O$_2$     & Methyl Hydroperoxide     & JANAF, \citet{Dorofeeva2001} \\
C$_1$O$_1$          & Carbon Monoxide          & JANAF, \citet{NIST} \\
C$_1$O$_2$          & Carbon Dioxide           & JANAF, \citet{NIST} \\
C$_1$Si$_1$         & Silicon Carbide          & JANAF, \citet{NIST} \\
C$_1$Si$_2$         & Silicon Carbide          & JANAF, \citet{NIST} \\
C$_2$               & Carbon                   & JANAF, \citet{NIST} \\
C$_2$H$_1$          & Ethynyl                  & JANAF, \citet{NIST} \\
C$_2$H$_2$          & Ethyne                   & JANAF, \citet{NIST} \\
C$_2$H$_2$O$_2$     & Glyoxal                  & JANAF, \citet{Dorofeeva2001} \\
C$_2$H$_2$O$_4$     & Oxalic Acid              & JANAF, \citet{Dorofeeva2001} \\
C$_2$H$_4$          & Ethene                   & JANAF, \citet{NIST} \\
C$_2$H$_4$O$_1$     & Oxirane                  & JANAF, \citet{NIST} \\
C$_2$H$_4$O$_3$     & Glycolic Acid            & JANAF, \citet{Dorofeeva2001} \\
C$_2$H$_6$O$_2$     & Dimethyl Peroxide        & JANAF, \citet{Dorofeeva2001} \\
C$_2$Si$_1$         & Silicon Carbide          & JANAF, \citet{NIST} \\
C$_2$Si$_2$         & Disilicarbide            & \citet{Tsuji1973} \\
C$_2$O$_1$          & CCO Radical              & JANAF, \citet{NIST} \\
C$_3$               & Carbon                   & JANAF, \citet{NIST} \\
C$_3$H$_1$          & 2-Propynylid\`{e}ne      & \citet{Tsuji1973} \\
C$_3$O$_2$          & Carbon Suboxide          & JANAF, \citet{NIST} \\
C$_4$               & Carbon                   & JANAF, \citet{NIST} \\
C$_4$H$_6$O$_4$     & Diacetyl Peroxide        & JANAF,\citet{Dorofeeva2001} \\
C$_5$               & Carbon                   & JANAF, \citet{NIST} \\
C$_1^+$             & Carbon + Ion               & JANAF, \citet{NIST} \\
C$_1^-$             & Carbon - Ion               & JANAF, \citet{NIST} \\
C$_1$H$_1^+$        & Methylidyne + Ion          & JANAF, \citet{NIST} \\
C$_1$H$_1^-$        & Methylidyne - Ion          & \citet{Tsuji1973} \\
C$_1$H$_1$O$_1^+$   & Formyl + Ion               & JANAF, \citet{NIST} \\
C$_1$O$_2^-$        & Carbon Dioxide - Ion       & JANAF, \citet{NIST} \\
C$_2^-$             & Carbon - Ion               & JANAF, \citet{NIST} \\
H$_1$Mg$_1$         & Magnesium Hydride        & \citet{BarklemCollet2016} \\
H$_1$Mg$_1$O$_1$    & Magnesium Hydroxide      & JANAF, \citet{NIST} \\
H$_1$O$_1$          & Hydroxyl                 & JANAF, \citet{NIST} \\
H$_1$O$_2$          & Hydroperoxo              & JANAF, \citet{NIST} \\
H$_1$Si$_1$         & Silylidyne               & JANAF, \citet{NIST} \\
H$_2$               & Hydrogen                 & JANAF, \citet{NIST} \\
H$_2$Mg$_1$O$_2$    & Magnesium Hydroxide      & JANAF, \citet{NIST} \\
H$_2$O$_1$          & Water                    & JANAF, \citet{NIST} \\
H$_2$O$_2$          & Hydrogen Peroxide        & JANAF, \citet{Dorofeeva2003} \\
H$_2$Si$_1$         & Silylene                 & \citet{Tsuji1973} \\
H$_3$Si$_1$         & Silyl Radical            & \citet{Tsuji1973} \\
H$_4$Si$_1$         & Silane                   & JANAF, \citet{NIST} \\
Mg$_1$O$_1$         & Magnesium Oxide          & JANAF, \citet{NIST} \\
Mg$_2$              & Magnesium                & JANAF, \citet{NIST} \\
O$_1$Si$_1$         & Silicon Oxide            & JANAF, \citet{NIST} \\
O$_2$               & Oxygen                   & JANAF, \citet{NIST} \\
O$_2$Si$_1$         & Silicon Oxide            & JANAF, \citet{NIST} \\
O$_3$               & Ozone                    & JANAF, \citet{NIST} \\
H$_1^+$             & Hydrogen + Ion             & JANAF, \citet{NIST} \\
H$_1^-$             & Hydrogen - Ion             & JANAF, \citet{NIST} \\
H$_1$Mg$_1$O$_1^+$  & Magnesium Hydroxide + Ion  & JANAF, \citet{NIST} \\
H$_1$O$_1^+$        & Hydroxyl + Ion             & JANAF, \citet{NIST} \\
H$_1$O$_1^-$        & Hydroxyl - Ion             & JANAF, \citet{NIST} \\
H$_1$Si$_1^+$       & Silylidyne + Ion           & JANAF, \citet{NIST} \\
H$_2^+$             & Hydrogen + Ion             & JANAF, \citet{NIST} \\
H$_2^-$             & Hydrogen - Ion             & JANAF, \citet{NIST} \\
H$_3$O$_1^+$        & Hydronium + Ion            & JANAF, \citet{NIST} \\
Mg$_1^+$            & Magnesium + Ion            & JANAF, \citet{NIST} \\
O$_1^+$             & Oxygen + Ion               & JANAF, \citet{NIST} \\
O$_1^-$             & Oxygen - Ion               & JANAF, \citet{NIST} \\
O$_2^+$             & Oxygen + Ion               & JANAF, \citet{NIST} \\
O$_2^-$             & Oxygen - Ion               & JANAF, \citet{NIST} \\
Si$_1^+$            & Silicon + Ion              & JANAF, \citet{NIST} \\
Si$_1^-$            & Silicon - Ion              & JANAF, \citet{NIST} \\
\hline
\caption{Gas phase species included in our simulations, with their chemical names and sources of thermochemical data. These species comprise all species bearing combinations of Mg, Si, C, O, and/or H in FASTCHEM COND \citep{FastchemCond}.}
\end{longtable}

\begin{table*}[h]
\centering
\begin{tabular}{ccccc}
\hline
\textbf{Condensate Species} & \textbf{Name} & \textbf{FASTCHEM Max. T (K)} & \textbf{Source} \\
\hline
C(s)               & Graphite                         & 6000 &  JANAF, \cite{NIST} \\
SiO$_2$(s,l)          & Silicon Dioxide                & 3000 & \citet{Barin1993} \\
Si(s,l)            & Silicon                        & 4500 & JANAF, \citet{NIST} \\
SiO(s)$^*$            & Silicon Oxide                  & 6000 & \citet{Gail2013SiO} \\
SiC(s)             & Silicon Carbide, Beta          & 4000 & JANAF, \citet{NIST} \\
MgH$_2$(s)         & Magnesium Hydride              & 2000 & JANAF, \citet{NIST} \\
Mg(OH)$_2$(s)      & Magnesium Hydroxide            & 1000 & JANAF, \citet{NIST} \\
Mg(s,l)            & Magnesium                      & 2000 & JANAF, \citet{NIST} \\
MgO(s,l)           & Magnesium Oxide, Periclase     & 5000 & JANAF, \citet{NIST} \\
MgSiO$_3$(s,l)     & Magnesium Silicate, Enstatite  & 3000 & JANAF, \citet{NIST} \\
Mg$_2$SiO$_4$(s,l) & Magnesium Silicate, Forsterite & 4000 & JANAF, \citet{NIST} \\
Mg$_2$Si(s,l)      & Magnesium Silicide             & 4000 & JANAF, \citet{NIST} \\
MgCO$_3$(s)        & Magnesium Carbonate            & 1000 & JANAF, \citet{NIST} \\
MgC$_2$(s)         & Magnesium Carbide              & 2500 & JANAF, \citet{NIST} \\
Mg$_2$C$_3$(s)     & Magnesium Carbide              & 2500 & JANAF, \citet{NIST} \\
\hline
\end{tabular}
\caption{Condensate species included in our simulations. These comprise all species bearing combinations of Mg, Si, C, O, and/or H in FASTCHEM COND \citep{FastchemCond}. The maximum temperature listed corresponds to either the triple point or the highest tabulated value \citep{FastchemCond}. We extrapolate all values to all conditions; the highest temperature we model is 4000~K. $^*$SiO(s) is not included in the canonical models -- the effects of its inclusion are discussed in Sec.~\ref{sec:speciation}.}
\end{table*}
\end{document}